\documentclass[aip,jcp,amsmath,amssymb,floatfix,reprint,citeautoscript,articletitle=false,noeprint]{revtex4-2}

\usepackage{amssymb}
\usepackage{amsmath}
\usepackage{graphicx}
\usepackage{dcolumn}
\usepackage{xcolor}
\usepackage{fancyhdr}
\usepackage[colorlinks,allcolors=black,citecolor=blue,urlcolor=blue]{hyperref}

\newcommand{\hlink}[1]{\textcolor{black}{#1}}

\begin{document}

\title{How Good Is DFT for Solid--Liquid Interfaces? A Comparison With the Random-Phase Approximation for Water on Graphene}

\author{Xavier R. Advincula}
\affiliation{Yusuf Hamied Department of Chemistry, University of Cambridge, Lensfield Road, Cambridge, CB2 1EW, UK}
\affiliation{Cavendish Laboratory, Department of Physics, University of Cambridge, Cambridge, CB3 0HE, UK}
\affiliation{Lennard-Jones Centre, University of Cambridge, Trinity Ln, Cambridge, CB2 1TN, UK}

\author{Yair Litman}
\affiliation{Max Planck Institute for Polymer Research, Ackermannweg 10, 55128 Mainz, Germany}
\affiliation{Max Planck Initiative for Liquid Research, Ackermannweg 10, 55128 Mainz, Germany}

\author{Jiuyang Shi}
\affiliation{Yusuf Hamied Department of Chemistry, University of Cambridge, Lensfield Road, Cambridge, CB2 1EW, UK}
\affiliation{Lennard-Jones Centre, University of Cambridge, Trinity Ln, Cambridge, CB2 1TN, UK}

\author{Flaviano Della Pia}
\affiliation{Yusuf Hamied Department of Chemistry, University of Cambridge, Lensfield Road, Cambridge, CB2 1EW, UK}
\affiliation{Department of Physics, Universitat Polit\`ecnica de Catalunya, 08019 Barcelona, Spain}

\author{Christoph Schran}
\email{cs2121@cam.ac.uk}
\affiliation{Cavendish Laboratory, Department of Physics, University of Cambridge, Cambridge, CB3 0HE, UK}
\affiliation{Lennard-Jones Centre, University of Cambridge, Trinity Ln, Cambridge, CB2 1TN, UK}

\author{Angelos Michaelides}
\email{am452@cam.ac.uk}
\affiliation{Yusuf Hamied Department of Chemistry, University of Cambridge, Lensfield Road, Cambridge, CB2 1EW, UK}
\affiliation{Lennard-Jones Centre, University of Cambridge, Trinity Ln, Cambridge, CB2 1TN, UK}

\begin{abstract}

Simulations of water at solid interfaces routinely disagree on basic properties, from contact angle to friction, depending on the model used, whether an empirical force field or a density functional theory (DFT) exchange-correlation (XC) approximation.
Yet no accurate computational reference has been available to determine which, if any, is reliable.
Here we assess XC approximations spanning the rungs of Jacob's Ladder against the random-phase approximation (RPA) at the graphene--water interface.
Using machine-learned potentials (MLPs) trained on both DFT and RPA data, we achieve the extensive sampling required to characterize interfacial structure, wettability, friction, and vibrational sum-frequency generation spectra.
We find that XC approximations disagree substantially with RPA and with each other.
To make sense of this, we introduce a transferable, multi-observable scoring framework that distills each functional's agreement with RPA into a single measure.
Six functionals emerge as reliable starting points for studies of graphene--water and other dispersion-dominated interfaces, namely B3LYP-D3(0), revPBE-D3(0), revPBE-D3(BJ), B97M-rV, r$^{2}$SCAN, and revPBE0-D3(0).
Notably, this ranking follows no simple rule based on functional family or rung of Jacob's Ladder.
What does hold across functionals is that dynamical and spectroscopic properties tend to be harder for DFT to reproduce than structural ones, with direct consequences for functional selection.
Our framework also offers a general strategy for benchmarking any atomistic model, including the new generation of foundational MLPs, against high-accuracy references at solid--liquid interfaces.
Beyond simulation, this work identifies which models can be trusted to connect measured contact angles, friction, and vibrational spectra to the molecular structure behind them.

\end{abstract}

\maketitle

\section*{Introduction}

Wolfgang Pauli remarked that ``God made the bulk; surfaces were invented by the devil'', to which we add: ``and the devil's big brother invented wet surfaces.''
We say this because a wet surface combines the reduced coordination of any surface~\cite{shen_science_1994, wat_inter_2016, nagata_chemrev_2020} with the complexity of liquid water itself~\cite{angelos_dft_water_2016}, whose structure emerges from a dynamic interplay of hydrogen bonding and van der Waals dispersion between closed-shell molecules.
This hydrogen-bond network rearranges on picosecond timescales~\cite{laage_2006_hynes}, and depending on the surface, water may adsorb strongly~\cite{carrasco_2012, limmer_metal_2013}, react and dissociate~\cite{feibelman_science_2002, advincula_reactivity_2026}, or interact only weakly through physisorption~\cite{ma_prb_2011, beate_2011, brandenburg_2019}.

Wet surfaces, that is, solid--liquid interfaces, underlie phenomena ranging from catalysis and energy storage to membrane separation and nanofluidic transport~\cite{wat_inter_2016,wat_charged_2021, lyderic_2010_rev}.
In all of these, macroscopic behavior, whether wetting, friction, or ion transport, is ultimately set by the molecular-scale structure and dynamics of the first few layers of interfacial water~\cite{limmer_metal_2013,tocci_friction,tocci_pnas_2024}.
Countless simulation studies have sought to capture this behavior, providing considerable insight~\cite{wat_inter_2016, striolo_2016_review, cicero_water_2008, ming_2016_natmat, flt_nanotubes}, and near-first-principles simulation of such complex, heterogeneous systems is now becoming routine~\cite{bartok_sciadv_2017, gabor_2019, unke_2021, complex_2021, pablo_mlps_water_2024}.
Accuracy, however, remains a significant frontier, raising the question of how far these simulations can be trusted.
When two distinct types of interaction compete, as hydrogen bonding and dispersion do here, the level of theory required to capture one may not be adequate for the other, and it is often unclear which approach, if any, is reliable for a given property.

Among such interfaces, graphene--water is a particularly important model system.
Its extreme surface-to-volume ratio has made graphene and related two-dimensional materials leading candidates for next-generation desalination and nanofluidic membranes~\cite{Wang2017_nanotech_pot, Xie2018_natnano, Fan2026_natwat}.
At the same time, its atomically flat, chemically inert surface isolates the water--water and solid--water interactions of interest from the added complexity of a heterogeneous surface chemistry.
How those two interactions balance to determine graphene's own wettability and transport behavior, however, is itself an open question.
Indeed, even for this seemingly simple case, experiments disagree substantially on how graphene behaves as a substrate for water.
Reported contact angles span a wide range depending on the substrate, layer number, surface preparation, and even airborne hydrocarbon contamination~\cite{Rafiee2012, Li2013_airbone, netz_contact_angle_2024, yair_wetting_2026, lim_2026}.
Separating graphene's intrinsic behavior from these external factors requires a molecular-level description of the interface whose accuracy has been established rather than assumed.

Atomistic simulation is the natural tool for this purpose.
Empirical force fields remain the most widely used approach, while density functional theory (DFT) is the highest level of theory in routine use, owing to its favorable balance of accuracy and cost~\cite{striolo_2016_review}.
However, different XC functionals can give substantially different pictures of water's structure and dynamics, a sensitivity documented independently in bulk water~\cite{angelos_dft_water_2016, klein_2021_bulk_water, cecilia_pnas_2022, pablo_dellago_2024, bingqing_2026_bulk_water, pablo_jcp_dynamics_2026} and at the air--water interface~\cite{yuki_dft_free_oh_2019}.
Despite this, no study has yet benchmarked DFT functionals consistently across structural, dynamical, and spectroscopic properties at a solid--liquid interface against a common, high-accuracy reference.
This is not a concern for simulation alone, since macroscopic measurements at these interfaces are increasingly interpreted through the molecular picture that simulation provides.

Such a benchmark is especially needed because, at a solid--liquid interface, a functional must capture both the water--water interactions responsible for hydrogen bonding and the solid--water interactions that govern adsorption and wetting, a balance that bulk water does not require.
At the graphene--water interface, this solid--water contribution is dominated by van der Waals (vdW) dispersion~\cite{ma_prb_2011, beate_2011}.
Standard GGAs and hybrids describe dispersion poorly, so it is usually added back through empirical corrections or non-local correlation~\cite{klimes_2012_review}.
Neither route is guaranteed to transfer to a system outside its parameterization.
As a result, even functionals sharing a nominally similar treatment of dispersion need not perform equally well at the interface.
Nor is the problem simply solved by climbing Perdew's Jacob's Ladder~\cite{perdew_ladder_2001}, which can improve either water--water or solid--water interactions on their own.
A functional that captures both faithfully, and for the right physical reasons, remains elusive.

A natural way to resolve this ambiguity is to benchmark DFT directly against a higher-tier electronic structure method.
Although many such methods exist, the options narrow considerably at solid--liquid interfaces, which generally require periodic simulations of around 100 atoms or more.
Higher-tier wave-function methods such as coupled cluster with single, double, and perturbative triple excitations [CCSD(T)] remain challenging to apply to extended, periodic systems of this size~\cite{gruneis_2019_ccsdt_mats}.
Of the various approaches, the random-phase approximation (RPA) is an appealing choice.
It captures the long-range electron correlation missing from semi-local and hybrid functionals, and treats dispersion and hydrogen bonding on the same footing without empirical parameters.
Its accuracy has been established across precisely the interactions at play here, from the relative stability of ice polymorphs~\cite{macher_rpa_ice_2014} to the relative densities of ice and liquid water~\cite{yao_rpa_2021}, a balance governed sensitively by the interplay of vdW and hydrogen bonding.
Most relevant to the present work, RPA reproduces the adsorption energy of a single water molecule on graphene to within a few meV of diffusion Monte Carlo and periodic CCSD(T) references~\cite{brandenburg_2019}.
Yet routine application of RPA in condensed-phase simulations has remained rare.
The reason is its steep computational scaling, typically between $\mathcal{O}(N^4)$ and $\mathcal{O}(N^5)$.
Reduced-scaling variants exist~\cite{wilhelm_2016_cubic}, but the cost is still prohibitive for the long, extended, periodic simulations required here.
Machine-learned potentials (MLPs) offer a way around this bottleneck.
By providing a direct mapping between atomic structure and energy, they bypass the need to solve the electronic structure problem at every step, enabling large-scale molecular dynamics at a fraction of the cost while retaining near first-principles accuracy~\cite{bartok_sciadv_2017, gabor_2019, pablo_mlps_water_2024}.
Combined with RPA, this approach has recently been extended beyond pure water to the more complex structure of aqueous electrolyte solutions~\cite{non_2024}.

Here, we use RPA as a reference to systematically assess a broad set of DFT XC functionals, spanning GGAs, meta-GGAs, and hybrids, at the graphene--water interface.
Using MLPs trained on both DFT and RPA reference data, we compare functional performance across interfacial water structuring and wettability, friction, and vibrational sum-frequency generation spectra, properties that determine the interfacial behavior of water probed experimentally.
We further introduce a transferable, multi-observable scoring framework for benchmarks at solid--liquid interfaces, designed with future comparisons in mind, whether against higher levels of electronic structure theory or at other interfaces.
The framework consolidates these comparisons into a single overall score for each functional, allowing us to identify which XC approximations most faithfully reproduce RPA-level accuracy.
All results are also available through an interactive website (see \hlink{Data Availability}), where the contribution of each observable to the overall score can be adjusted and every score recomputed accordingly.
Our aim throughout is to establish the accuracy of the molecular description that connects the microscopic structure of the interface to the macroscopic behavior measured in experiment.

\begin{figure*}[htp!]
    \centering
    \includegraphics[width=\textwidth]{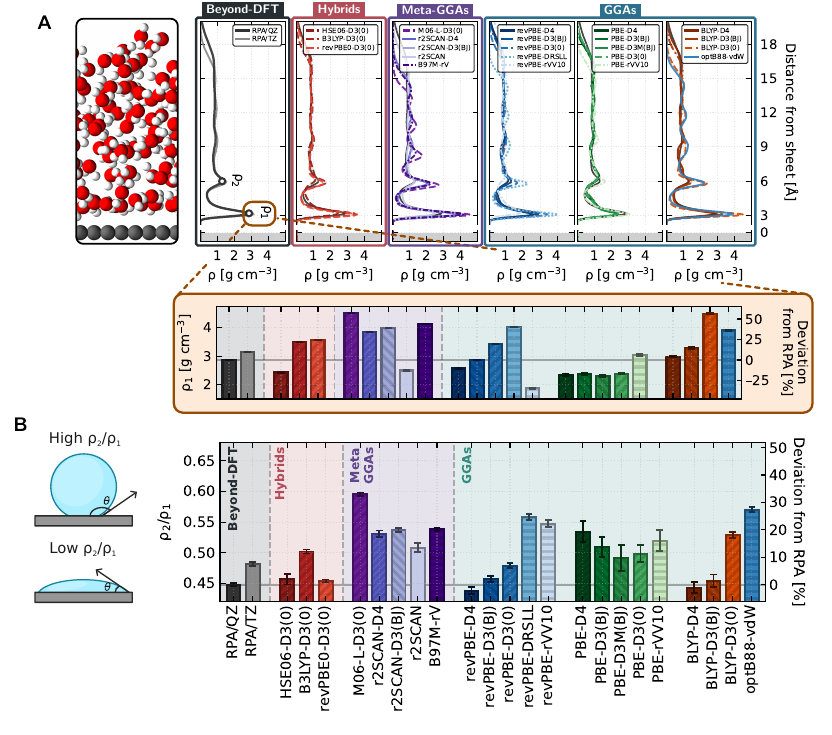}
    \caption{\textbf{Big impact of the XC approximation on interfacial water structuring and wettability at the graphene--water interface.}
    (A) Water oxygen density profiles as a function of the distance from the graphene sheet, computed with RPA/QZ (our reference) and RPA/TZ, and grouped by XC family: hybrids, meta-GGAs, and GGAs, each paired with a variety of dispersion corrections.
    Below the main panel the first interfacial peak density, $\rho_1$, is shown for each functional, alongside its percentage deviation from the RPA reference.
    (B) Ratio of the second to first interfacial peak densities, $\rho_2/\rho_1$, used as a structural proxy for wettability, for each XC approximation.
    Percentage deviations from the RPA reference are shown alongside, with the RPA reference indicated by the gray horizontal line.
    }
    \label{fig:fig1}
\end{figure*}

\section*{Results}

In this work, we compare the performance of various DFT XC functionals with the many-body method RPA.
The observables we focus on require long simulation times that are generally out of reach for direct first-principles calculations.
We therefore represent each level of theory with an MLP trained on that method's reference data, a strategy now well established for comparing the structural and dynamical properties of water across functionals~\cite{pablo_dellago_2024, pablo_jcp_dynamics_2026, hubert_2026_convergence, oneill2026_liq_revpbe, bingqing_2026_bulk_water}.
Each training set combines a common basis dataset, recomputed at every level of theory, with theory-specific structures sampled from the corresponding potential energy surface, so that each MLP targets its own reference method (see \hlink{Methods}).
The resulting MLPs allow us to accumulate several nanoseconds of simulation for every level of theory considered.
Each MLP has been rigorously benchmarked against its reference method (see \hlink{Sections S1 and S2}), so we treat its predictions as representative of that method and refer to each simply by the method's name throughout.

There is, of course, an endless list of DFT functionals we could test.
We selected a subset that samples across the rungs of Jacob's Ladder, with each successive rung incorporating additional physical ingredients into the XC description.
In particular, we considered the GGA functionals PBE~\cite{revpbed3_1}, BLYP~\cite{blyp_1_1988, lyp_1988}, revPBE~\cite{revpbed3_1,comment_ggas_1998}, and optB88-vdW~\cite{klimes_2010_optb88}; the meta-GGA functionals M06-L~\cite{nonlocal_m06_2006}, r$^{2}$SCAN~\cite{r2scan_2020}, and B97M-rV~\cite{b97m_2015}; and the hybrid functionals B3LYP~\cite{blyp_1_1988, lyp_1988, becke_exact_1993}, revPBE0~\cite{revpbed3_1,comment_ggas_1998,pbe0_1999}, and HSE06~\cite{screened_hybrid_2003,screen_hybrid_2_2006}, each paired with a range of dispersion corrections reflecting commonly used pairings in the literature.
Functionals combining GGA exchange with nonlocal van der Waals correlation~\cite{drsll_2004}, such as optB88-vdW and the DRSLL- and rVV10-paired variants, are grouped with the GGAs throughout, as is common practice.
Going beyond DFT, we use RPA as our reference level of theory.
RPA correlation energies converge slowly with basis-set size, so that a triple-$\zeta$ (TZ) basis still carries appreciable incompleteness error, and a quadruple-$\zeta$ (QZ) basis is needed for converged results~\cite{kresse_rpa_mlff_2023}.
Our reference is therefore RPA/QZ, which, given the steep computational scaling associated with larger basis sets, we obtain through a delta-learning approach, training on the difference between RPA/QZ and RPA/TZ rather than on RPA/QZ data directly (see \hlink{Methods}).
We note that our benchmark is against RPA rather than against experiment by design.
Several of the observables we consider are not directly accessible to experiment, and for those that are, reported measurements vary substantially, as the spread in measured contact angles illustrates~\cite{Rafiee2012, Li2013_airbone, netz_contact_angle_2024, yair_wetting_2026, lim_2026}.
A disagreement with experiment could therefore reflect that variability as much as the XC approximation itself, whereas comparing against RPA isolates the latter.

\begin{figure*}[htp!]
    \centering
    \includegraphics[width=0.88\textwidth]{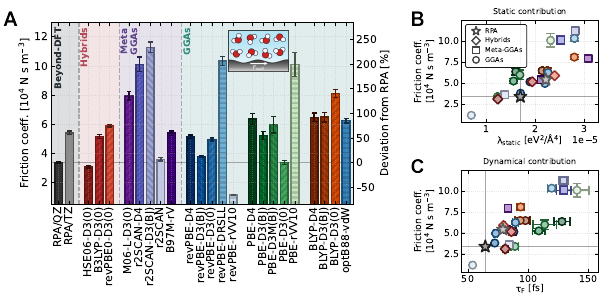}
    \caption{\textbf{Big impact of the XC approximation on the friction coefficient at the graphene--water interface.}
    (A) Friction coefficients computed with RPA/QZ (our reference) and RPA/TZ, and grouped by XC family: hybrids, meta-GGAs, and GGAs, each paired with a variety of dispersion corrections.
    Percentage deviations from the RPA reference are shown alongside, with the RPA reference indicated by the gray horizontal line.
    (B, C) Decomposition of the friction coefficient into its static and dynamical contributions, plotted against (B) $\lambda_{\textrm{static}}$, a measure of the corrugation of the graphene--water energy landscape, and (C) $\tau_\textrm{F}$, the force correlation time capturing the relaxation of lateral forces at the interface.
    Colors follow (A) throughout, and solid lines mark the RPA reference along both axes.
    }
    \label{fig:fig2}
\end{figure*}

\subsection*{How is water structured at graphene? Interfacial layering and wettability}

In Figure~\ref{fig:fig1}A, we assess the ability of these methods to predict the solid--liquid structuring of water on rigid graphene walls, as captured by the water density profiles (for a comparison with flexible walls, see \hlink{Section S3}).
Across all methods, these profiles reveal the same qualitative picture, with water clearly stratified into distinct layers near the graphene sheet, consistent with previous observations of water at solid surfaces~\cite{cicero_water_2008, tocci_friction, Ruiz-Barragan_JPCL_2019, tocci_pnas_2024, advincula_inter_nano_2026}.
Beneath this shared qualitative agreement, however, the height and width of each peak, and hence the degree of interfacial structuring, vary considerably from one method to the next.
Among the meta-GGAs, for instance, M06-L-D3(0) predicts substantially more pronounced structuring than r$^2$SCAN.
This disagreement is most pronounced for the first density peak, $\rho_{1}$, corresponding to the water layer in direct contact with graphene.
DFT functionals not only differ substantially from one another, but span roughly $\pm$45\% relative to the RPA reference, a spread far larger than one would expect for a seemingly simple structural quantity.
This variation highlights how sensitive the predicted graphene--water interface is to the choice of XC approximation.

A useful structural proxy for wettability is the ratio of the second to first interfacial peak densities, $\rho_2/\rho_1$.
Liquid-state structural order near a surface is known to control nanoscale wetting~\cite{tanaka_linet_2026}, and the relative height of successive density layers offers a direct measure of that order.
Indeed, a direct, monotonic correlation has been shown to emerge between this ratio and the macroscopic contact angle, with a higher $\rho_2/\rho_1$ corresponding to a larger contact angle (more hydrophobic behavior), and a lower ratio to a smaller one (more hydrophilic behavior)~\cite{Agosta2026}.
In Figure~\ref{fig:fig1}B, we show this quantity, where we again observe substantial variability across DFT XC functionals, spanning roughly $-$10\% to $+$30\% relative to the RPA reference.
Notably, this variability is not symmetric.
DFT functionals tend to predict higher $\rho_2/\rho_1$ ratios than RPA, and therefore graphene as more hydrophobic than RPA suggests, although this ratio alone cannot distinguish a weaker graphene--water attraction from an overly strong water--water one.
Reassuringly, revPBE-D3(0), used in a recent study of the strain dependent wettability of graphene~\cite{lim_2026}, is in reasonable agreement with RPA for this ratio.
The sensitivity of $\rho_2/\rho_1$ is not limited to the choice of functional either.
Within a single family, changing only the dispersion correction shifts both the $\rho_1$ contact-layer density and the $\rho_2/\rho_1$ ratio, as the revPBE functionals show~\cite{behler_damping_2025}.

\begin{figure*}[htp!]
    \centering
    \includegraphics[width=\textwidth]{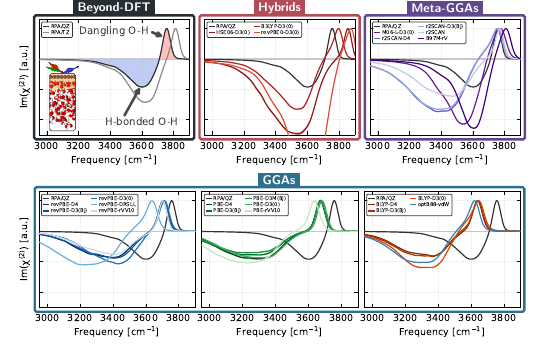}
    \caption{\textbf{Big impact of the XC approximation on the VSFG spectrum of the graphene--water interface.}
    Theoretical $\mathrm{Im}(\chi^{(2)})$ spectra computed with RPA/QZ (our reference) and RPA/TZ and grouped by XC family, namely hybrids, meta-GGAs, and GGAs, each paired with a variety of dispersion corrections, using the ssVVCF methodology~\cite{ssvcf_2015}.
    The RPA/QZ spectrum is shown in every panel for reference, and all spectra are normalized to their maximum value.
    The inset in the RPA panel shows a schematic of the system considered, highlighting the topmost interfacial water layer, whose dangling and H-bonded O-H groups dominate the spectral response (see \hlink{Methods}).
    }
    \label{fig:fig3}
\end{figure*}

\subsection*{How slippery is graphene? Friction and its static and dynamical origins}

Moving beyond a purely structural analysis, we now examine how the choice of DFT functional influences an experimentally relevant dynamical property, the friction between the solid and the liquid, which governs interfacial transport.
Here, we consider flexible graphene walls, whose surface fluctuations couple to the local curvature and to fluid-solid interactions at the interface, and can thereby influence flow~\cite{ming_2016_natmat, flt_nanotubes, ke_zhou_2025_flex, lim_2026} (for a comparison with the rigid-wall case, see \hlink{Section S3}).
Friction coefficients were computed from equilibrium MD simulations using the Green--Kubo relation~\cite{bocquet_barrat_friction_1994, bocquet_barrat_2014}, fitting the running integral to an analytical expression~\cite{joly_analytical_2019} to circumvent the well-known ``plateau'' problem in finite systems (see \hlink{Methods}).

From Figure~\ref{fig:fig2}, we see that once again the different DFT functionals exhibit a wide range of variability in the friction coefficient, ranging from roughly $-$65\% to $+$230\% relative to the RPA reference.
Such a large variation points to a subtle balance of interfacial and dynamical effects underlying friction, which is highly sensitive to the choice of XC approximation.
Notably, the friction coefficient is markedly harder to converge for the PBE family than for the other functionals, consistent with an overstructured, sluggish hydrogen-bond network~\cite{angelos_dft_water_2016}.

To understand the origin of this variability, we decompose the friction coefficient into its static and dynamical contributions, $\lambda_{\textrm{static}}$ and $\tau_{\textrm{F}}$, respectively (Figure~\ref{fig:fig2}B)~\cite{tocci_static_dyn_2020}.
The static contribution, $\lambda_{\textrm{static}}$, can be related to the free energy barrier for the lateral motion of the oxygen atom of a water molecule in the contact layer, and is hence a measure of the corrugation of the graphene--water energy landscape.
The dynamical contribution, $\tau_{\textrm{F}}$, is instead the correlation time of the lateral force acting on the fluid, so that a higher $\tau_{\textrm{F}}$ corresponds to a slower relaxation of the lateral forces at the interface.
Comparing functionals with RPA for each contribution, we find that both scatter substantially, but not in the same way.
Most functionals overestimate $\lambda_{\textrm{static}}$, although around a third fall below the RPA value.
For $\tau_{\textrm{F}}$, by contrast, all but one functional overestimate it, placing RPA at the low end of the range.
Lateral forces at the interface therefore decorrelate faster in RPA than in almost any of the DFT descriptions.
This is consistent with the less pronounced interfacial structuring RPA predicts, seen in its comparatively low $\rho_2/\rho_1$ ratio, and with the sluggish dynamics of the overstructured PBE family noted above.
Together, these observations suggest that functionals which overstructure the contact layer also slow its relaxation.
Beyond this systematic offset, the scatter along both axes indicates that different functionals miss the reference for different underlying reasons, with no single mechanism responsible.
Repeating this decomposition with rigid graphene walls (see \hlink{Section S3}) leaves the scatter in $\lambda_{\textrm{static}}$ largely unchanged, so this scatter originates in the description of the interface itself rather than in the fluctuations of the sheet.
Wall flexibility does, however, systematically lower the absolute friction coefficient.
The spread we observe across functionals at this single interface is also comparable, for both $\lambda_{\textrm{static}}$ and $\tau_{\textrm{F}}$, to that reported by Tocci \textit{et al.}~\cite{tocci_static_dyn_2020} across entirely different two-dimensional materials, where both quantities vary substantially between graphene and MoS$_2$.
In other words, an inadequate choice of XC functional can introduce as much spread as switching to a different two-dimensional material altogether.
This is an important reminder that friction predictions are only as trustworthy as the electronic structure description behind them.

It is instructive to place these numbers in the context of experiments.
Our friction coefficients lie in the same range as previous machine-learned simulations of water on flat carbon~\cite{flt_nanotubes, tocci_pnas_2024, ke_zhou_2025_flex}, and within the range inferred from measurement~\cite{Xie2018_natnano, chen_slip_2022} (see \hlink{Section S3}).
The reported experimental values are widely scattered, however, varying by more than an order of magnitude across techniques and samples.
This spread has been attributed to the surface charge of the graphene and its interaction with the supporting substrate, rather than to the intrinsic interface~\cite{Xie2018_natnano, li_ming_2022}.
Measurements on graphite capillaries~\cite{boya_2021_natcomms_friction} are harder still to compare against.
For water on graphite, electronic contributions to friction, absent from any classical description, have been proposed as a significant term~\cite{kavokine_friction_2022, bui_classical_quantum_2023, sutter2026_rev}.
Simulation and experiment are therefore not measuring quite the same quantity.
What we report is the friction of a clean interface, whereas a measurement also includes whatever the substrate and surface chemistry of the particular sample add, which is what makes this comparison hard.

\begin{figure*}[htp!]
    \centering
    \includegraphics[width=0.85\textwidth]{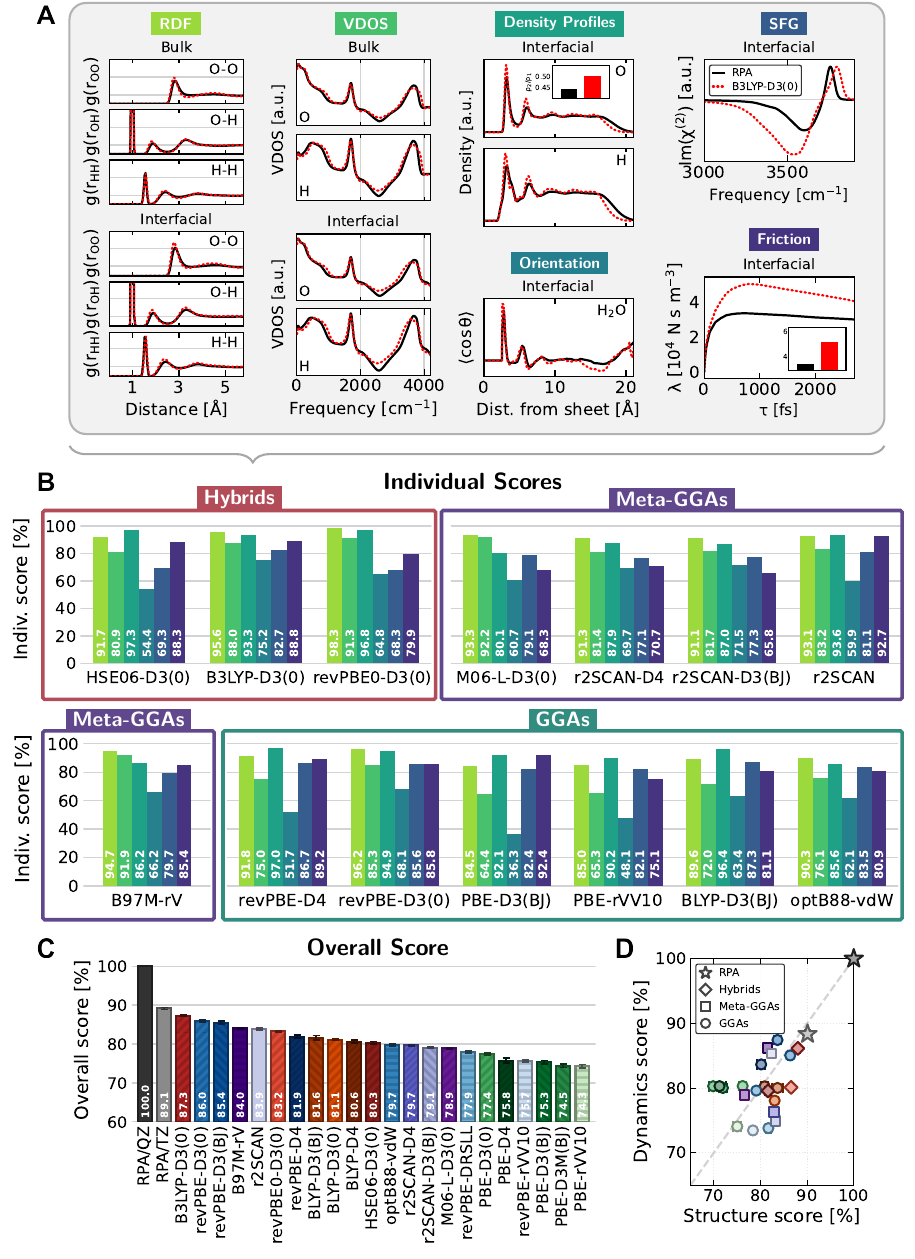}
    \caption{\textbf{Identifying the most suitable XC approximation through a multi-observable scoring framework.}
    (A) Overview of the six observables entering the scoring framework: radial distribution functions (RDF), vibrational density of states (VDOS), density profiles, orientational order, VSFG spectra, and friction, shown for RPA/QZ and B3LYP-D3(0) as an illustrative comparison (see Sections~S4 and~S5 for further details).
    (B) Individual scores for representative XC approximations, grouped by family (hybrids, meta-GGAs, and GGAs) and decomposed by observable, with colors matching the observable labels in (A).
    (C) Overall score for each XC approximation, obtained by averaging over all individual observable scores.
    (D) Dynamics score (averaged over the VDOS, VSFG, and friction scores) plotted against structure score (averaged over the RDF, density profile, and orientation scores), for all XC approximations considered, with colors matching those in (C).
    }
    \label{fig:fig4}
\end{figure*}

\subsection*{Can vibrational spectra tell functionals apart? Interfacial signatures and comparison with experiment}

A particularly valuable experimentally accessible probe of interfacial water structure is vibrational sum-frequency generation (VSFG) spectroscopy~\cite{shen_og_1989_nature, shen_science_1994, saykally_sfg_2018, wang_ange_2024, ions_gra_advin_2025}.
Because VSFG is a second-order nonlinear optical process, it is forbidden in centrosymmetric media such as bulk liquid water, but becomes allowed at interfaces, where inversion symmetry is broken.
This makes it an intrinsically surface-specific probe.
The measured signal is directly proportional to the second-order nonlinear susceptibility, $\chi^{(2)}$, which is resonantly enhanced when the input infrared frequency matches a vibrational transition of interfacial water molecules, yielding a vibrationally resolved spectrum of the interface.
The sign of the resonant, imaginary part of this susceptibility, $\mathrm{Im}(\chi^{(2)})$, in turn reflects the net orientation of water molecules at the interface.
A positive signal arises from O-H bonds pointing away from the bulk liquid and toward the interface, as for a dangling O-H group.
A negative signal arises from O-H bonds pointing into the bulk, as is typical of hydrogen-bonded O-H groups.
This orientational picture holds strictly only for water molecules whose O-H stretches are effectively decoupled from one another, as is the case for a dangling O-H group~\cite{tahara_2008_up_down,nagata_chemrev_2020}.
Extracting this signal from simulation is considerably more challenging than computing bulk vibrational spectra, since only the handful of molecular layers at the interface contribute, rather than the entire system.
As a result, reliably resolving the interfacial features requires extensive statistical sampling~\cite{ssvcf_2015, morita_book, yair_sfg_2023} (see \hlink{Methods}).
Given this sampling cost, we restrict our analysis to rigid graphene walls, which lets us isolate the impact of the XC approximation on interfacial water structure without the extra degree of freedom introduced by wall fluctuations.

In Figure~\ref{fig:fig3}, we show the VSFG spectra of the graphene--water interface computed with different DFT functionals and with RPA/QZ (our reference) and RPA/TZ, with the dangling O-H and hydrogen-bonded O-H regions labeled in the beyond-DFT panel.
All spectra exhibit a sharp positive peak at 3600-3900~cm$^{-1}$, arising from the dangling O-H group of interfacial water molecules, and a broad negative peak centered at 3200-3600~cm$^{-1}$, arising from hydrogen-bonded O-H groups.
Inspection of the spectra reveals noticeable differences across XC functionals.
Since all spectra are normalized to their maximum value (the dangling O-H peak), these differences are most apparent in the hydrogen-bonded region.
Relative to RPA, the meta-GGA r$^{2}$SCAN-based functionals produce a broader negative feature whose minimum is shifted to lower frequency~\cite{sauer_2026_dft_zeolites}.
The same is true of the GGAs (see e.g. the PBE-based functionals), so this behavior is not specific to r$^{2}$SCAN.
A hydrogen-bonded band shifted to lower frequency is the usual signature of overly strong hydrogen bonds, which weaken the covalent O-H bond and lower its stretching frequency, consistent with the known tendency of semilocal functionals to overbind water~\cite{angelos_dft_water_2016}.
The hybrids behave differently, with minima close to the RPA value but a sharper and deeper feature.
Their admixture of exact (Hartree-Fock) exchange partly corrects the self-interaction error of semilocal functionals~\cite{angelos_dft_water_2016}.
This strengthens the covalent O-H bond and correspondingly weakens the hydrogen bonds, bringing the position of the band back toward RPA.
The dangling O-H peak also shifts between functional families.
Relative to RPA, the GGAs lower its frequency by roughly 85~cm$^{-1}$ on average, while the meta-GGAs raise it by roughly 15~cm$^{-1}$.
Because this O-H group is not hydrogen-bonded, its frequency reflects only the intrinsic stiffness of the O-H bond predicted by each functional.
This makes it a useful check on why the hydrogen-bonded band appears at lower frequency, which could stem either from an intrinsically softer O-H bond or from overly strong hydrogen bonding.
For the meta-GGAs, the dangling O-H peak sits at slightly higher frequency than in RPA, so their O-H bond is, if anything, stiffer.
The lower frequency of their hydrogen-bonded band must therefore reflect overly strong hydrogen bonding.
Notably, all functionals predict a redshift of the dangling O-H frequency at the graphene--water interface relative to the air--water interface (see \hlink{Section S3}).
This indicates a weak but non-negligible interaction between the dangling O-H group and the graphene sheet~\cite{ohto_gra_hbn_2018, wang_ange_2024}.
We note that although revPBE0-D3(0) still yields a reasonable redshift between the two interfaces, the shape of its air--water spectrum differs from that previously reported with the same functional~\cite{yuki_dft_free_oh_2019}.
This likely reflects differences in electronic structure settings, to which simulations of water are known to be sensitive~\cite{hubert_2026_convergence, oneill2026_liq_revpbe}, as ours follow those of Ref.~\citenum{bingqing_2026_bulk_water} (see \hlink{Section S1}).

Having compared the functionals against RPA, we now turn to experiment.
The graphene--water to air--water redshift is the quantity best suited for this comparison.
Heterodyne-detected SFG measurements have been reported for a freely suspended graphene--water interface~\cite{wang_ange_2024}, which, like our simulations, is an isolated monolayer free of substrate effects.
Absolute peak positions, by contrast, are less useful for this purpose.
Our simulations treat the nuclei classically and thus omit nuclear quantum effects, which substantially lower the frequency of a mode as anharmonic as the O-H stretch.
The computed frequencies therefore come out systematically too high, whereas taking the difference between two interfaces computed with the same method largely cancels this offset.
The measured redshift is $30 \pm 10$~cm$^{-1}$~\cite{wang_ange_2024}.
An earlier measurement on the same system reported a larger value~\cite{xu_tian_2023}, which the more recent work~\cite{wang_ange_2024} attributes to a bulk contribution and a graphene background that it does not itself observe.
Across the functionals we test, the redshift ranges from $22$ to $94$~cm$^{-1}$ (see \hlink{Section S3}).
Most GGAs and meta-GGAs overestimate it, the largest value being roughly three times the measured one.
Only five of the methods considered fall within the experimental uncertainty, namely RPA/QZ, HSE06-D3(0), B3LYP-D3(0), revPBE0-D3(0), and revPBE-D3(0).

\subsection*{Which functionals should we trust? A multi-observable scoring framework}

A central goal of this study is to identify which DFT functional best reproduces the RPA reference across the full range of properties considered.
Such a comparison can help reassess conclusions drawn from previous DFT-based studies of solid--liquid interfaces, and establish the level of electronic structure theory required for a reliable description.
It can also build confidence in DFT and machine-learned approaches that model these systems accurately at a fraction of the cost of RPA, which remains largely intractable at scale (see \hlink{Section S2}).

To this end, we introduce a transferable, multi-observable scoring framework that evaluates each functional against several properties relevant to solid--liquid interfaces.
It provides an efficient and robust measure of each functional's overall agreement with RPA.
The observables entering this framework, shown in Figure~\ref{fig:fig4}A, are the radial distribution functions (RDFs), vibrational density of states (VDOS), density profiles (including the wetting proxy $\rho_2/\rho_1$), orientational profiles, VSFG spectra, and friction (including both the friction coefficient and the running Green--Kubo integral).
For each observable, we compute a score measuring how far a given level of theory deviates from the RPA reference.
This is based either on the difference between the corresponding functions (e.g., for RDFs and VDOS) or on that between scalar values (e.g., for the friction coefficient)~\cite{complex_2021}.
Full details are provided in \hlink{Section S5}.
These scores are then averaged within each category to yield the individual scores shown in Figure~\ref{fig:fig4}B.
For B3LYP-D3(0), as illustrated, the scores obtained from the bulk and interfacial O-O, O-H, and H-H RDFs are averaged to give an overall RDF score of 95.6\%, indicating close agreement with the RPA/QZ reference.
The same procedure yields individual scores for each of the remaining categories.
Having access to these individual scores, rather than a single aggregate number, is valuable in its own right.
In many applications one may only be interested in a functional's ability to reproduce a particular structural, dynamical, or spectroscopic property.
Comparing across categories, the orientational profile is the harshest structural test and VSFG, its dynamical counterpart, the harshest dynamical one.
A detailed breakdown of the individual scores for all DFT functionals considered is provided in \hlink{Section S3}.

Averaging the six individual scores in turn yields an overall score for each functional (Figure~\ref{fig:fig4}C), providing a single, interpretable metric for how faithfully each XC approximation reproduces RPA across all properties considered.
This average weights all six observables equally, a choice that can be varied on the accompanying website (see \hlink{Data Availability}), where every score is recomputed accordingly.
Among the functionals examined, B3LYP-D3(0), revPBE-D3(0), and revPBE-D3(BJ) perform best, closely followed by B97M-rV, r$^{2}$SCAN, and revPBE0-D3(0), with all six scoring above 83\%.
This ranking, obtained entirely against a theoretical reference, finds independent support in experiment.
Three of the six, B3LYP-D3(0), revPBE0-D3(0), and revPBE-D3(0), are also among the five methods that reproduce the measured redshift of the dangling O-H peak to within its experimental uncertainty (see \hlink{Section S3}).
This comparison played no part in constructing the scores.
By contrast, the ranking is not anticipated by the static benchmark most commonly used to assess functionals at such interfaces.
The adsorption energy of a single water molecule on graphene shows no significant correlation with the overall score in any of the three adsorption configurations considered (see \hlink{Section S3}).
This highlights the importance of benchmarking functionals against the thermodynamic and dynamical properties that emerge from long-time sampling, rather than relying on static binding energies alone.

Notably, RPA/TZ itself scores markedly lower than RPA/QZ across all observables, a consequence of basis-set incompleteness.
This error is far more severe for methods such as RPA, which rely on unoccupied (virtual) orbitals, than for DFT, where only the occupied orbitals enter (see \hlink{Section S2}).
Finally, grouping the individual scores into a dynamics-related score (averaged over the VDOS, VSFG, and friction scores) and a structure-related score (averaged over the RDF, density profile, and orientation scores) allows us to compare how well each functional reproduces structural versus dynamical properties (Figure~\ref{fig:fig4}D).
Slightly over half of the functionals fall below the diagonal in this representation, indicating that a faithful description of interfacial dynamics tends to be harder to achieve than one of interfacial structure.
This is an important finding for the future development and benchmarking of XC approximations.

\section*{Discussion}

We assessed a broad set of DFT XC functionals at the graphene--water interface, using MLPs to access the structural, dynamical, and spectroscopic properties that govern this interface.
As our reference, we chose a beyond-DFT method, RPA, which performs well when assessed against experimental measurements as well as against diffusion Monte Carlo adsorption energies for a single water molecule on graphene~\cite{brandenburg_2019}.
We find that DFT functional performance varies substantially, with sizable deviations from RPA in both interfacial contact density and friction.
Importantly, this variation does not reduce to a simple rule such as functional family or rung of Jacob's Ladder.
Even within a single XC functional family, changing only the dispersion correction can substantially shift both quantities.
Encouragingly, six functionals, B3LYP-D3(0), revPBE-D3(0), revPBE-D3(BJ), B97M-rV, r$^{2}$SCAN, and revPBE0-D3(0), reproduce RPA closely across all observables considered.
They offer a practical, well-justified starting point for future DFT-based studies of the graphene--water interface and, more broadly, of nonpolar solid--water systems where dispersion-dominated adsorption plays a similarly central role.
Their precise ordering, however, is not unique.
The overall score weights the six observables equally, and the ranking among the leading functionals can shift once that choice is changed.
A study concerned primarily with friction, for instance, has no reason to weight the RDFs as heavily.
To make this choice explicit rather than implicit, we provide an interactive version of these results (see \hlink{Data Availability}), where the weight of each observable, and the reference itself, can be set freely and every score recomputed accordingly.
Reassuringly, the top three functionals reported here, namely B3LYP-D3(0), revPBE-D3(0), and revPBE-D3(BJ), remain consistently near the top under the range of alternative weightings we tested.

Central to this effort was the use of MLPs, which made the necessary structural and dynamical sampling accessible at every DFT and RPA level of theory considered, sampling that would otherwise have been out of reach.
This was especially critical for VSFG spectroscopy, where resolving the subtle interfacial signal requires far more extensive statistical sampling than bulk spectroscopies demand~\cite{ssvcf_2015, morita_book, yair_sfg_2023}.
More importantly, this sampling takes the assessment beyond the static benchmarks traditionally used to evaluate XC functionals at interfaces, which is essential for a comprehensive view of their performance.
A static benchmark would in fact yield a quite different ranking, since ordering the functionals by the adsorption energy of a single water molecule on graphene does not reproduce the one obtained here (see \hlink{Section S3}).
How well a functional describes an isolated molecule is therefore not a reliable guide to how it describes the interface at finite temperature, and benchmarks at solid--liquid interfaces should move toward finite-temperature observables wherever possible.

We also note that achieving this level of accuracy required careful attention to RPA itself.
Our choice of RPA/QZ, rather than the more affordable RPA/TZ, as our reference was not incidental.
The smaller TZ basis suffers from basis-set incompleteness errors large enough to noticeably lower its scores relative to RPA/QZ across all observables.
Indeed, RPA/TZ offers only a marginal improvement over the best hybrid functionals, so the added cost of an RPA calculation at this basis-set size buys little of the accuracy that motivates going beyond DFT in the first place.
This is a useful reminder that a beyond-DFT reference is only as reliable as the numerical settings underlying it, which should therefore be checked carefully.
The concern is not specific to RPA, since published simulations of liquid water with the same functional can disagree substantially depending on the code and numerical settings used, particularly when those settings are not fully converged~\cite{hubert_2026_convergence, oneill2026_liq_revpbe}.
Such variability arises when results from different codes and workflows are compared, and is largely removed here by construction, since the same code and workflow were used for every level of theory considered.
Residual convergence errors will of course remain, and will differ between methods.
Our scores are therefore best read as relative to RPA/QZ under a common protocol, which is the comparison they are intended to support.

In applying our multi-observable scoring framework, we found that dynamical and spectroscopic properties tend to be harder for DFT to reproduce than purely structural ones, with slightly over half of the functionals falling below the diagonal when the two are compared directly.
This matters in practice, since friction and VSFG spectra, which belong to this harder group, are also the two properties most directly comparable to experiment.
A functional selected on the basis of its structural agreement with a high-level reference is therefore not guaranteed to perform equally well against experimentally measured dynamical or spectroscopic observables, and should be validated against these properties independently.
In practice, however, this is easier said than done.
Reported friction coefficients on carbon surfaces scatter more widely than the functionals do, since what is measured depends heavily on the substrate, surface charge, and preparation of the sample.
In spectroscopy, comparatively small differences in sample preparation and analysis have likewise been enough to change the reported picture of the same interface.
No single measurement can therefore settle the question on its own, which is part of why we have scored functionals against a common theoretical reference rather than against any one experiment.

We expect the scoring framework presented here, and the protocol underlying it, to be of use well beyond the present study.
It makes no assumption specific to a given level of theory, nor to the graphene--water interface itself.
In principle, it can therefore be used to assess any model capable of generating the required trajectories and spectra against a common, high-accuracy reference, whether RPA/QZ as used here or, where available, experimental measurements directly.
This makes it well suited both to the wide variety of empirical classical force fields in use at these interfaces and to the new generation of foundational MLPs~\cite{Deng2023_chgnet, mace_mp0_2025, Yuan2026_foundational}, whose transferability to a specific interface is rarely tested against the properties that matter there.
To this end, we release the scoring pipeline as a standalone package (see \hlink{Data Availability}), which takes a trajectory from any model and returns its score against the reference for each observable.
This places new potentials on the same footing as the functionals assessed here.

To the best of our knowledge this is the first ``beyond-DFT'' simulation study of a solid liquid interface. 
Nonetheless, a natural next step is to extend this approach to still higher levels of theory.
Like any approximation, RPA itself has known systematic errors, and in particular tends to underbind weakly interacting systems owing to missing single-excitation contributions and the neglect of exchange effects beyond the ring approximation~\cite{ren_2011_excitations, ren_2012_review, eshuis_2012_methods}.
Recent progress in coupling machine-learned potentials with CCSD(T) for condensed-phase systems~\cite{prl_ccsdt_2022,niamh_ccsdt_water_2025,oneill_2026_caco,nore_2026_aromatic_ccsdt} suggests that the strategy we have applied here with RPA may soon be extended to that level of theory, provided the periodic nature of the graphene sheet can be captured.
A CCSD(T) reference would open the way to a direct comparison with experiment rather than between levels of theory alone.
Nuclear quantum effects would then also have to be included, since they become a limiting source of error once quantitative agreement with experiment is the aim.
Even without these extensions, this work shows that functional choice at solid--liquid interfaces should rest not on precedent or convenience, but on evidence from the interface itself.

\section*{Methods}

\textbf{Machine Learning Potentials.}
For all the MLPs developed in this work we used the MACE architecture~\cite{mace_canonical}, with 128 equivariant channels, two message-passing layers, and a 6~{\AA} cutoff.
The resulting models capture semi-local interactions through an effective receptive field, which extends to 12~{\AA}, corresponding to the product of the number of layers and the cutoff distance per layer. 
All electronic structure calculations used to generate the training data were carried out with the CP2K/Quickstep code~\cite{cp2k_2020}.
Each MLP was trained following a two-generation iterative improvement protocol.
A first generation establishes a common basis dataset, adapted from a dataset originally developed in Ref.~\citenum{advincula_reactivity_2026} and recomputed at each level of electronic structure theory, followed by a second generation that adds theory-specific structures sampled from the corresponding potential energy surface.
The dataset spans bulk water, the air--water interface, and water at graphene, both in mono- and bilayer confinement and at the open interface, with the sheet held frozen or flexible.
Full details of the training procedure are provided in \hlink{Section S2}.
For the RPA/QZ model specifically, we adopted a $\Delta$-learning strategy: a baseline MLP is trained on RPA/TZ data, and a second $\Delta$-MLP is trained on top of it to capture the energy and force differences between RPA/QZ and RPA/TZ.
In this way, the final RPA/QZ prediction is obtained by summing the outputs of the two models.
Additional details are provided in \hlink{Section S2}.

\textbf{Molecular Dynamics Simulations.}
All simulations were performed using the ASE software~\cite{ase_2017} with the MLPs developed in this work.
For each level of theory, production results were obtained from 12 independent runs, each comprising a 60~ps equilibration period followed by 250~ps of production sampling.
The temperature was controlled to 300~K in the NVT ensemble using a CSVR thermostat~\cite{csvr_2007} with a 0.05~ps time constant.
The dynamics were propagated classically, with a 0.5~fs time step using standard hydrogen masses.
All systems were modeled in orthorhombic simulation cells with periodic boundary conditions applied in all three spatial directions.
To prevent spurious interactions between periodic replicas, we introduced a 15~{\AA} vacuum along the axis perpendicular to the interface.
This exceeds the model's effective receptive field, fully decoupling periodic images along that direction.
In all cases, graphene sheets, either frozen or flexible, with lateral dimensions of 17.290~{\AA}~$\times$~17.112~{\AA}, were used.
All uncertainties reported throughout, whether shown as error bars or as shaded regions, correspond to the standard error across these 12 runs, propagated as appropriate for the specific quantity considered.

\textbf{Data Analysis.}
Friction coefficients were calculated from equilibrium MD simulations using the Green--Kubo relation~\cite{bocquet_barrat_friction_1994, bocquet_barrat_2014, joly_analytical_2019}, with the total force on all solid atoms recorded every 1~fs along both in-plane dimensions.
Vibrational sum-frequency generation spectra were computed using the ssVVCF approach~\cite{ssvcf_2015}.
We specifically focused on the topmost interfacial water layer, which dominates the spectral response as shown in our previous work~\cite{ions_gra_advin_2025}.
Additional details are provided in \hlink{Section~S4}.

\begin{acknowledgements}

A.M. is indebted to Peijun Hu for lighting the spark of scientific curiosity in his life and then subsequent decades of mentorship and friendship.
X.R.A. and A.M. acknowledge support from the European Union under the ``n-AQUA'' European Research Council project (Grant No. 101071937).
C.S. acknowledges financial support from the Isaac
Newton Trust G122390 and the UKRI reference EP/V062654/1.
This work used the ARCHER2 UK National Supercomputing Service via the UK’s HEC Materials Chemistry Consortium, funded by EPSRC (EP/F067496).
We also utilized resources from the Cambridge Service for Data Driven Discovery (CSD3), supported by EPSRC (EP/T022159/1) and DiRAC funding, with additional access through a University of Cambridge EPSRC Core Equipment Award (EP/X034712/1).
We further acknowledge EuroHPC Joint Undertaking for awarding the project ID EHPC-REG-2024R02-130 access to Leonardo at CINECA (Italy), and project ID EHPC-REG-2025R02-112 access to JUPITER at J{\"u}lich (Germany).

\end{acknowledgements}

\section*{Data Availability}

All data required to reproduce the findings of this work will be made openly available on GitHub upon acceptance of this manuscript.
An interactive version of the scores reported here, in which the observable weights and the reference can be varied, is available at \url{https://water-ice-group.github.io/water-graphene-benchmark/}.

\section*{References}
%

%
%
%
%
%
%
%

%
\end{document}


\def\mytitle{How Good Is DFT for Solid--Liquid Interfaces? A Comparison With the Random-Phase Approximation for Water on Graphene}

\title{Supporting Information for: \mytitle}

\author{Xavier R. Advincula}
\affiliation{Yusuf Hamied Department of Chemistry, University of Cambridge, Lensfield Road, Cambridge, CB2 1EW, UK}
\affiliation{Cavendish Laboratory, Department of Physics, University of Cambridge, Cambridge, CB3 0HE, UK}
\affiliation{Lennard-Jones Centre, University of Cambridge, Trinity Ln, Cambridge, CB2 1TN, UK}

\author{Yair Litman}
\affiliation{Max Planck Institute for Polymer Research, Ackermannweg 10, 55128 Mainz, Germany}
\affiliation{Max Planck Initiative for Liquid Research, Ackermannweg 10, 55128 Mainz, Germany}

\author{Jiuyang Shi}
\affiliation{Yusuf Hamied Department of Chemistry, University of Cambridge, Lensfield Road, Cambridge, CB2 1EW, UK}
\affiliation{Lennard-Jones Centre, University of Cambridge, Trinity Ln, Cambridge, CB2 1TN, UK}

\author{Flaviano Della Pia}
\affiliation{Yusuf Hamied Department of Chemistry, University of Cambridge, Lensfield Road, Cambridge, CB2 1EW, UK}
\affiliation{Department of Physics, Universitat Polit\`ecnica de Catalunya, 08019 Barcelona, Spain}

\author{Christoph Schran}
\email{cs2121@cam.ac.uk}
\affiliation{Cavendish Laboratory, Department of Physics, University of Cambridge, Cambridge, CB3 0HE, UK}
\affiliation{Lennard-Jones Centre, University of Cambridge, Trinity Ln, Cambridge, CB2 1TN, UK}

\author{Angelos Michaelides}
\email{am452@cam.ac.uk}
\affiliation{Yusuf Hamied Department of Chemistry, University of Cambridge, Lensfield Road, Cambridge, CB2 1EW, UK}
\affiliation{Lennard-Jones Centre, University of Cambridge, Trinity Ln, Cambridge, CB2 1TN, UK}

\keywords{}

{\maketitle}
\tableofcontents

\onecolumngrid

\FloatBarrier

\newpage

\section{Electronic Structure} \label{sec:elect_settings}
\subsection{DFT settings}

All DFT calculations were performed using the CP2K/Quickstep code~\cite{cp2k_2020} to generate both total energies and atomic forces.
To sample the range of common exchange-correlation (XC) approximations, we selected functionals spanning several rungs of Jacob's ladder, from generalized-gradient approximations (GGAs) through meta-GGAs to hybrid functionals: the GGA functionals PBE~\cite{revpbed3_1}, BLYP~\cite{blyp_1_1988,lyp_1988}, revPBE~\cite{revpbed3_1,comment_ggas_1998}, and optB88-vdW~\cite{klimes_2010_optb88}; the meta-GGA functionals M06-L~\cite{nonlocal_m06_2006}, r$^2$SCAN~\cite{r2scan_2020}, and B97M-rV~\cite{b97m_2015}; and the hybrid functionals B3LYP~\cite{blyp_1_1988,lyp_1988,becke_exact_1993}, revPBE0~\cite{revpbed3_1,comment_ggas_1998,pbe0_1999}, and HSE06~\cite{screened_hybrid_2003, screen_hybrid_2_2006}.
As in the main text, functionals combining GGA exchange with nonlocal van der Waals correlation~\cite{drsll_2004}, such as optB88-vdW and the DRSLL- and rVV10-paired variants, are grouped with the GGAs throughout, as is common practice.
Each functional was paired with one or more dispersion corrections from the Grimme family, D3(0)~\cite{revpbed3_2}, D3(BJ)~\cite{bj_2011}, and D4~\cite{d4_1_2017, d4_2_2019}, as well as the nonlocal van der Waals (vdW) corrections DRSLL~\cite{drsll_2004} and rVV10~\cite{vv10_2010, rvv10_2013}.
In the case of optB88-vdW, only the DRSLL pairing was considered, since its exchange was reparameterised specifically for use with that kernel~\cite{klimes_2010_optb88}.
All hybrid functionals were evaluated using the auxiliary density matrix method (ADMM)~\cite{admm_2010} to reduce the cost of Hartree-Fock exchange, and for these only D3(0) was tested, given the higher cost that exchange involves.
The combinations were chosen to reflect commonly used pairings in the literature and to span a range of dispersion correction approaches.
Not all corrections were applied to every functional.

All calculations employed the mixed Gaussian and plane-wave (GPW) scheme~\cite{gpw_1997} as implemented in CP2K/Quickstep.
Core electrons were treated using Goedecker-Teter-Hutter (GTH) pseudopotentials~\cite{gth_1996}.
The Kohn-Sham valence orbitals of hydrogen and oxygen were expanded in a TZV2P-GTH basis set~\cite{jurg_2007_basis}, comprising triple-$\zeta$ valence Gaussians with two sets of polarization functions, while carbon was described using the DZVP-MOLOPT-SR-GTH basis set, comprising double-$\zeta$ valence Gaussians with one set of polarization functions~\cite{jurg_2007_basis}.
The electron density was expanded in an auxiliary plane-wave basis with a cutoff of 1050 Ry, using a five-level multi-grid with a relative cutoff of 60 Ry.
For hybrid functional calculations, the Hartree-Fock exchange was evaluated using ADMM\cite{admm_2010}, employing the \texttt{cpFIT3} fitting basis set. 
For the global hybrids B3LYP and revPBE0, Hartree-Fock exact exchange fractions of 20\% and 25\% respectively were used, evaluated with a truncated Coulomb operator-\cite{truncated_col_2009} and a cutoff radius of 6.0~\AA.
This setup for the hybrid functionals was inspired by that of Ref.~\citenum{bingqing_2026_bulk_water}.
For HSE06, a range-separated scheme was used instead, retaining 25\% short-range exact exchange with a range-separation parameter of $\omega$ = 0.11~bohr$^{-1}$.

\subsection{Random-Phase Approximation}

Random-phase approximation (RPA) total energies and atomic forces were computed using the CP2K/Quickstep code~\cite{cp2k_2020} within the mixed GPW formalism~\cite{gpw_1997}.
The RPA correlation energy and exact-exchange (EXX) energy were evaluated from Kohn-Sham eigenfunctions obtained with the PBE functional~\cite{revpbed3_1}, and the resolution-of-identity (RI) approximation~\cite{hutter_2023_ri} was used to reduce the cost of the four-center integrals.
Triple-zeta correlation-consistent basis sets were used for H and O, and the basis sets of Ye and Berkelbach~\cite{berkelbach_2022} were used for C.
Auxiliary basis sets for the RI integrals were taken from CP2K defaults for H and O, while those for C were generated following the automatic procedure of Stoychev et al.~\cite{aux_bss_2017,lehtola_2021}
GTH pseudopotentials~\cite{gth_1996} were used throughout.
A plane-wave cutoff of 1050 Ry was used for the electron density, with a separate cutoff of 300 Ry for the wavefunction correlation integrals, the latter of which has been shown to converge atomic forces to within 0.00001~meV/\AA~\cite{non_2024}.
The RPA correlation energy was evaluated on a 20-point minimax quadrature grid, introducing errors below 1~mHartree according to literature~\cite{quadrature_rpa_2015}.
As discussed in Section~\ref{sec:mlp_details}, non-negligible differences were observed between RPA energies and forces computed with triple-zeta (TZ) and quadruple-zeta (QZ) basis sets.
To reduce basis set incompleteness errors, single-point RPA energies were therefore computed on an additional set of smaller configurations using quadruple-zeta (QZ) basis sets for all species, and the resulting QZ$-$TZ correction was added to the TZ-level dataset to yield an effective RPA/QZ reference.
A detailed explanation of this procedure is given in Section~\ref{sec:mlp_details}.

\newpage
\section{Development and validation of machine learning potentials} \label{sec:mlp_details}

\subsection{Details of the Models}
For all the machine learning potentials (MLPs) models, we used the MACE architecture~\cite{mace_canonical}, with 128 equivariant channels, two message-passing layers, and a 6~{\AA} cutoff.
The resulting models capture semi-local interactions through an effective receptive field, which extends to 12~\AA, corresponding to the product of the number of layers and the cutoff distance per layer.

\subsection{Workflow and Validation of the DFT Models}

\begin{figure}
    \centering
    \includegraphics[width=\linewidth]{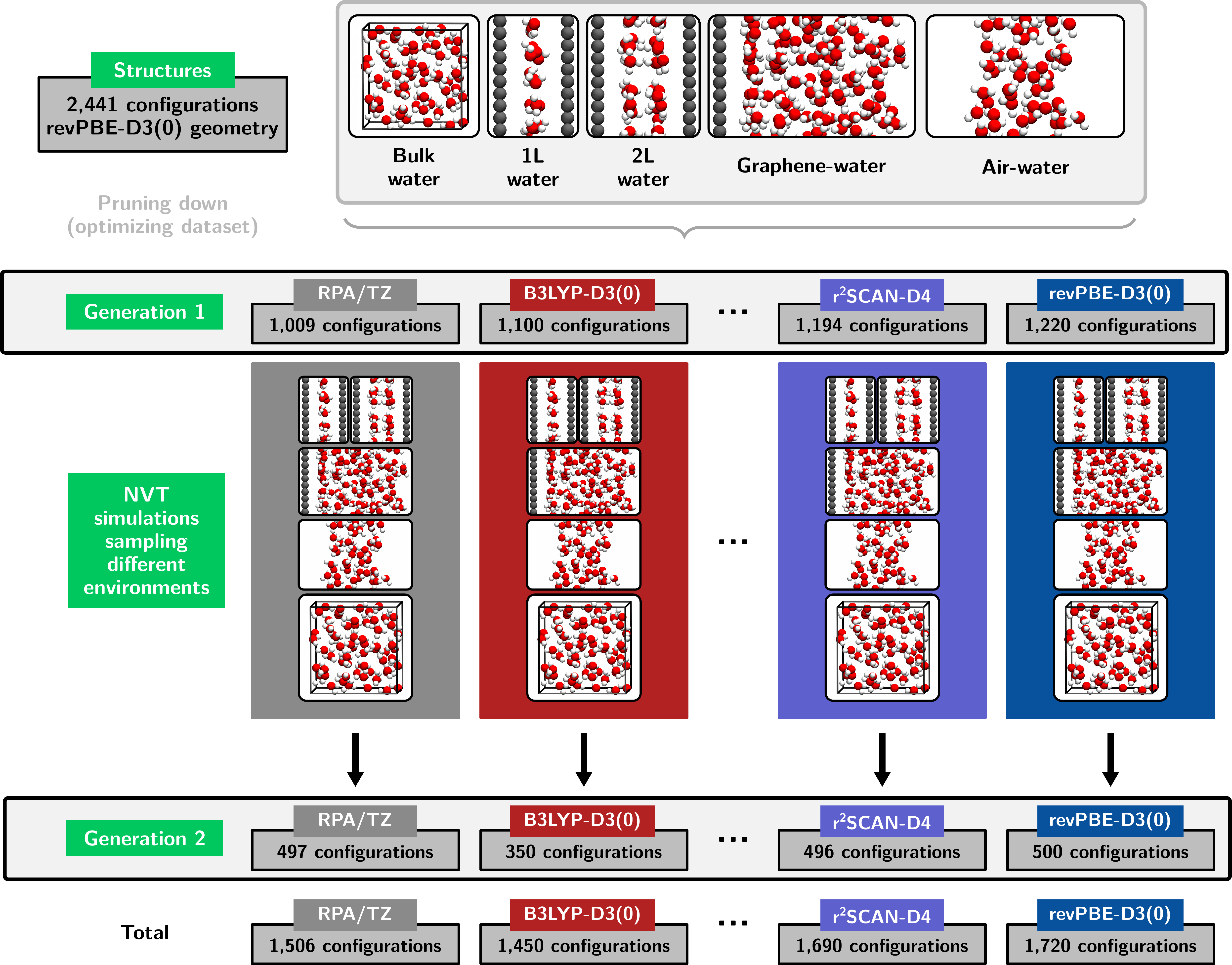}
    \caption{\textbf{Schematic of the training procedure for 4 representative ML models trained in this work.}
    The starting 2{,}441 structures with revPBE-D3(0) geometries were obtained from Ref.~\citenum{advincula_reactivity_2026}.}
    \label{fig:si_model_gens}
\end{figure}

We developed MLPs trained to describe interfacial water at the graphene--water interface at different levels of electronic structure theory, spanning several DFT functionals with distinct XC approximations.
The training of the models was divided into two generations, as graphically depicted in Figure~\ref{fig:si_model_gens}.
The first generation establishes a common basis dataset recomputed at each level of theory, while the second generation adds theory-specific structures sampled from the potential energy surface (PES) of each level of theory.

In Table~\ref{tab:rmse_all_structures} we report a summary of the force and energy root mean square error (RMSE) for each MLP model, together with the corresponding number of base (first-generation) and additional (second-generation) structures.

\begin{figure}
    \centering
    \includegraphics[width=\linewidth]{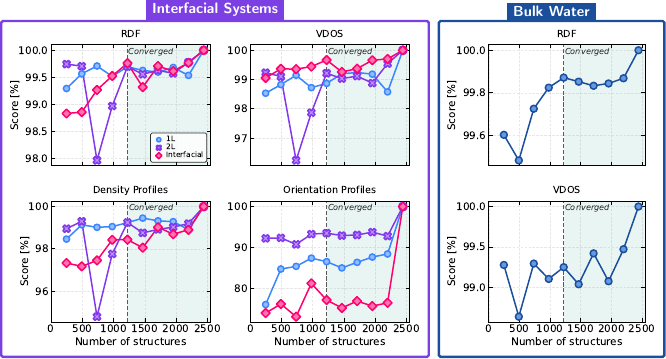}
    \caption{\textbf{Self-consistent convergence of the first-generation basis dataset.}
    Accuracy of MLPs trained on progressively smaller subsets of the full 2{,}441 revPBE-D3(0) structures, evaluated against the RDF, VDOS, density profile, and orientational profile scores defined in Section~\ref{sec:scoring}, for the three interfacial systems considered, monolayer confined water (1L), bilayer confined water (2L), and graphene--water interfacial water (Interfacial), as well as for bulk water.
    The scores are computed over the liquid species only, namely the O-O, O-H and H-H pairs for the RDFs and O and H for the VDOS and density profiles, in both the interfacial and bulk cases.
    The highlighted region marks dataset sizes of 1{,}220 structures and above, beyond which we consider the properties of interest to be converged.
    We therefore take 1{,}220 structures as our basis dataset.
    }
    \label{fig:si_self_consist_basis}
\end{figure}

\textbf{First generation.}
The first generation comprised a common training set for all models, adapted from previous work~\cite{advincula_reactivity_2026}, using geometries obtained at the revPBE-D3(0) level of theory.
From this starting point, we excluded configurations with excessively large lateral dimensions ($L_x$ or $L_y$ greater than 15~\AA{}), as these substantially increase computational cost, particularly when using high levels of electronic structure theory.
We further excluded carbon nanotube configurations, as well as ice configurations, since no ice conditions are sampled in this work.
Removing both allowed us to construct a more compact and targeted dataset of 2{,}441 structures, comprising monolayer confined water (1L), bilayer confined water (2L), graphene--water interfacial water, air--water interfacial water, and bulk water.
This number of structures, however, is too computationally demanding for methods such as RPA.
We therefore aim to identify a well-converged basis dataset.
That is, a set of geometries large enough to reliably capture the structural and dynamical properties of interest, but small enough to be recomputed at each level of electronic structure theory considered, including RPA.
To this end, we optimize the dataset size in a self-consistent manner, training models on progressively smaller subsets of the revPBE-D3(0) geometries and validating that a reduced dataset can achieve performance comparable to a model trained on the full set of 2{,}441 structures.
This is assessed using a subset of the scoring system described in Section~\ref{sec:scoring}, focusing on structural and dynamical properties that can be obtained without extensive simulation, since many models must be tested at this stage.
Specifically, we validate radial distribution functions (RDFs), vibrational density of states (VDOS), density profiles, and orientational profiles at the interface, as well as RDFs and VDOS in the bulk.
Properties requiring more extensive sampling, such as vibrational sum-frequency generation (VSFG) spectroscopy, are not used at this validation stage.
We consider dataset sizes ranging from 100\% down to 10\% of the full 2{,}441 structures, in steps of 10\%.
The results of this validation are shown in Figure~\ref{fig:si_self_consist_basis}.
Following this procedure, we identify 1{,}220 structures as a suitable basis dataset, as shown in Figure~\ref{fig:si_self_consist_basis}.
This basis dataset of revPBE-D3(0) geometries is then recomputed at each level of electronic structure theory considered, yielding the dataset used to train the first-generation models.
We note that, in a small percentage of cases, single-point calculations at a given level of theory fail to converge, and the corresponding geometries are simply excluded for that level of theory.

\begin{figure}
    \centering
    \includegraphics[width=\linewidth]{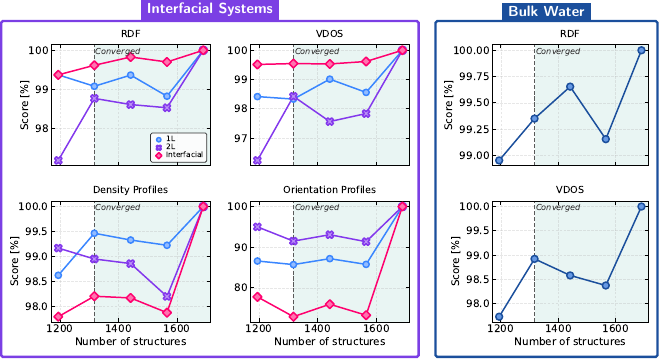}
    \caption{\textbf{Self-consistent convergence of the second-generation additional dataset for a representative example.}
    Accuracy of MLPs trained with progressively larger additional datasets, sampled from r$^2$SCAN trajectories, evaluated against the RDF, VDOS, density profile, and orientational profile scores defined in Section~\ref{sec:scoring}, for the three interfacial systems considered, monolayer confined water (1L), bilayer confined water (2L), and graphene--water interfacial water (Interfacial).
    The highlighted region marks the point beyond which adding further structures no longer significantly improves model accuracy.
    For r$^2$SCAN, this is roughly achieved at 1{,}317 total structures, corresponding to 123 structures added to the 1{,}194-structure basis dataset.
    We note that this validation is for test purposes only and, since all the data had already been computed, for the final r$^2$SCAN model we used the largest additional dataset available, 495 structures, corresponding to the last point shown at 1{,}689 total structures.
    }
    \label{fig:si_self_consist_additional_r2scan}
\end{figure}

\textbf{Second generation.}
The second generation expands this basis dataset with new configurations, sampled at each level of electronic structure theory using the corresponding first-generation model, so as to explore the PES specific to that level of theory beyond the revPBE-D3(0) geometries used in the first generation.
To this end, we ran 600~ps simulations at each level of theory, using a graphene sheet with lateral dimensions of 12.755~{\AA}~$\times$~12.173~{\AA}, across the five representative conditions: 1L, 2L, graphene--water interfacial water, air--water interfacial water, and bulk water, with the graphene sheet held either frozen or flexible in the cases involving a solid surface.
New configurations were then obtained by randomly subsampling structures from these trajectories.
As before, we determine a well-converged additional dataset in a self-consistent manner, checking that a small set of newly sampled structures already reproduces the structural and dynamical properties of interest as reliably as a larger set of such structures.
The results of this validation are shown for a representative example in Figure~\ref{fig:si_self_consist_additional_r2scan}.
As in the first generation, a small percentage of these additional structures fail to converge and are excluded.
For the final models, we use the largest number of additional structures available, since these had already been computed.
In most cases this amounts to 500 structures.
We emphasize, however, that the purpose of the validation above was to confirm that even smaller additional datasets already yield converged properties, giving us confidence that each MLP reliably reproduces the properties of interest at its underlying level of theory.

\subsection{Workflow and Validation of the RPA Model}
We developed an MLP trained to describe interfacial water at the graphene--water interface at the RPA level of theory, following the same two-generation procedure described above.
For the first generation, we recomputed the optimized basis dataset of  1{,}220 structures at the RPA/TZ level of theory.
As before, a small percentage of structures failed to converge and were excluded.
For the second generation, we sampled additional configurations at RPA/TZ using the corresponding first-generation model, so as to explore the RPA PES beyond the revPBE-D3(0) geometries used in the first generation, again following the self-consistent validation procedure described previously and excluding the small fraction of structures that failed to converge.

\begin{figure}
    \centering
    \includegraphics[width=0.75\linewidth]{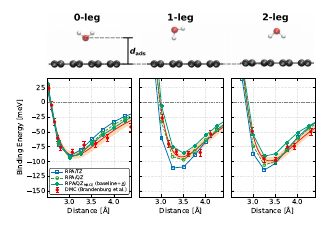}
    \vspace{-0.7cm}
    \caption{\textbf{Binding-energy curves for a single water molecule on graphene, shown for the 0-leg (left), 1-leg (middle), and 2-leg (right) adsorption motifs.}
    Side and top views of each motif are shown alongside the corresponding curve.
    The motifs and the DMC reference data are reproduced from Ref.~\citenum{brandenburg_2019}.
    RPA/QZ is necessary to reach agreement with DMC, highlighting the importance of basis set convergence for accurately describing carbon-water interactions.
    }
    \label{fig:si_rpa_binding_energy}
\end{figure}

\begin{figure}
    \centering
    \includegraphics[width=0.7\linewidth]{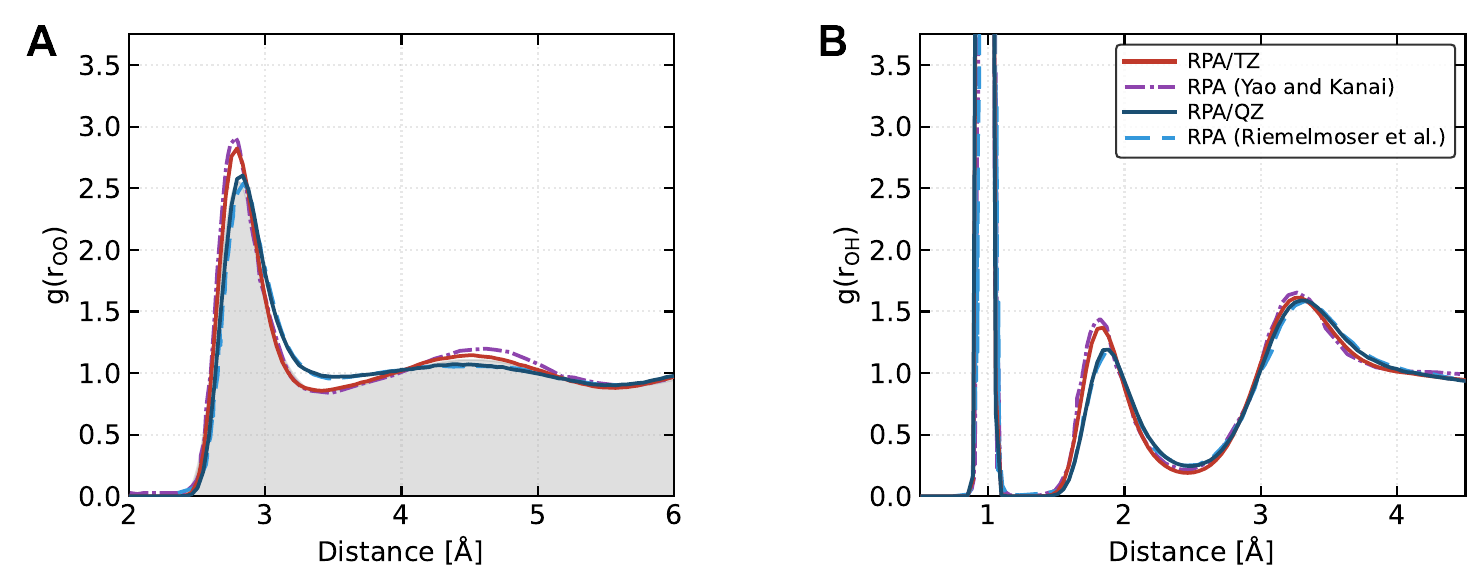}
    \vspace{-0.5cm}
    \caption{\textbf{Radial distribution functions of liquid water at ambient conditions ($T=300$~K, $\rho=1$~g/cm$^{3}$).}
    Details of the MD simulations are given in the Methods section.
    (A) Oxygen-oxygen and (B) oxygen-hydrogen radial distribution functions.
    RPA results from Riemelmoser et al.~\cite{kresse_rpa_mlff_2023} use a converged (QZ) basis set, while those from Yao and Kanai~\cite{yao_rpa_2021} use a TZ basis set.
    Experimental data are taken from Ref.~\citenum{skinner_2013} and are shown as a shaded gray region.
    Our RPA/TZ results agree with the Yao and Kanai reference~\cite{yao_rpa_2021}, but only once we move to RPA/QZ do we recover agreement with Riemelmoser et al.~\cite{kresse_rpa_mlff_2023} and with experiments~\cite{skinner_2013}, consistent with the basis set incompleteness errors suggested in Ref.~\citenum{kresse_rpa_mlff_2023}.
    }
    \label{fig:si_rpa_rdf}
\end{figure}

\textbf{From RPA/TZ to RPA/QZ.}
The RPA/TZ dataset built in this way is not yet our definitive reference, as RPA/TZ is known to suffer from basis set incompleteness errors~\cite{kresse_rpa_mlff_2023}.
In particular, Eshuis and Furche~\cite{eshuis_2012} showed that QZ quality basis sets are typically required to converge RPA calculations, and we confirm that this is also the case for our systems in Figures~\ref{fig:si_rpa_binding_energy} and~\ref{fig:si_rpa_rdf}, where RPA/TZ and RPA/QZ predictions differ.
To correct for this basis set incompleteness error, we adopt a $\Delta$-learning approach.
Computing an entire reference dataset directly at the RPA/QZ level would be prohibitively expensive given RPA's steep computational scaling with basis set size, so rather than training a new MLP directly on RPA/QZ data, we train a $\Delta$-MLP on the difference between RPA/QZ and RPA/TZ, and add its prediction on top of the RPA/TZ baseline MLP already obtained above.
Our final RPA/QZ description therefore relies on two MLPs: the RPA/TZ baseline model described above, and a RPA/TZ~$\rightarrow$~RPA/QZ $\Delta$-MLP trained on top of it.
This strategy follows the general idea that a machine-learned correction can be added to the predictions of a computationally cheap, low-level baseline method (here, RPA/TZ) to approximate a more expensive, high-level target method (here, RPA/QZ), 
\vspace{-0.4cm}
\begin{equation}
    y_{\textrm{target}} = y_{\textrm{baseline}} + \Delta^{\textrm{target}}_{\textrm{baseline}}.
\end{equation}
Equations~\ref{eq:rpa_total_e} and~\ref{eq:rpa_total_f} illustrate how this decomposition is applied to the energies and forces used in this work,
\begin{equation}
    E_{\textrm{RPA/QZ}} = E_{\textrm{RPA/TZ}} + E_{\textrm{RPA/QZ}-\textrm{RPA/TZ}}, \label{eq:rpa_total_e}
\end{equation}
\begin{equation}
    \mathbf{F}_{\textrm{RPA/QZ}} = \mathbf{F}_{\textrm{RPA/TZ}} + \mathbf{F}_{\textrm{RPA/QZ}-\textrm{RPA/TZ}}. \label{eq:rpa_total_f}
\end{equation}
We note that the $\Delta$-MLP is trained not only on energy differences but also on force differences.
This makes for a substantially more data-efficient approach than one relying on energies alone, since each configuration provides only a single energy value, whereas it provides $3N$ force components (three Cartesian components per atom), offering far richer information per configuration.

\textbf{Constructing the $\Delta$-MLP dataset.}
Because RPA/QZ calculations are substantially more expensive than RPA/TZ due to their unfavourable cost scaling, we build the $\Delta$-MLP training set as a small, targeted dataset rather than reusing the full first- and second-generation RPA/TZ dataset directly.
Specifically, we run additional NVT simulations using a model trained on the combined first- and second-generation RPA/TZ data, and extract configurations spanning bulk water and graphene--water interfacial systems at various coverages, including a small number of single-water-on-graphene configurations to accurately capture binding energies.
Due to the memory constraints associated with RPA/QZ calculations on larger systems, these configurations are kept comparatively small.
Therefore, we used bulk water boxes of 32 water molecules at experimental density, and graphene--water configurations with in-plane cell dimensions of $L_x \times L_y = 9.88 \times 8.556$~\AA$^2$.
The number of structures included in the $\Delta$-MLP dataset is reported in Table~\ref{tab:rmse_all_structures}.

\textbf{Validation.}
As shown in Figures~\ref{fig:si_rpa_binding_energy} and~\ref{fig:si_rpa_rdf}, this $\Delta$-learning approach successfully recovers an RPA/QZ-like description from the RPA/TZ baseline.
These results confirm that a TZ basis set alone is insufficient for RPA.
Only once the RPA/QZ correction is applied do we recover the RPA results previously reported by Brandenburg et al.~\cite{brandenburg_2019}, which are in agreement with diffusion Monte Carlo (DMC) values, one of the most accurate methods available, as well as the RDFs reported by Riemelmoser et al.~\cite{kresse_rpa_mlff_2023}, which are in good agreement with experiments.

\newpage

\phantom{...}

\newpage

\phantom{...}
\newpage

\phantom{...}
\newpage

\newpage
\phantom{table}

\begin{table}[h]
\centering
\caption{Summary of force and energy RMSE for each ML model, together with the corresponding number of base (first-generation) and additional (second-generation) structures.}
\label{tab:rmse_all_structures}
\vspace{0.4cm}
\begin{tabular}{lccccc}
\toprule
 & Base & Additional & Energy RMSE & Force RMSE \\
\multirow{-2}{*}{Model} & structures & structures & {[meV/atom]} & {[meV/\AA]} \\
\midrule
\multicolumn{5}{l}{\textit{GGAs}} \\
PBE-D4        & 1220 & 500 & 0.40 & 14.80 \\
PBE-D3(0)     & 1220 & 500 & 2.20 & 16.60 \\
PBE-D3(BJ)    & 1220 & 500 & 0.30 & 17.40 \\
PBE-D3M(BJ)   & 1220 & 500 & 0.30 & 15.80 \\
PBE-rVV10     & 1220 & 500 & 0.30 & 17.10 \\
BLYP-D4       & 1219 & 500  & 0.50 & 39.40 \\
BLYP-D3(0)    & 1220 & 500 & 1.70 & 40.30 \\
BLYP-D3(BJ)   & 1220 & 500  & 0.60 & 40.30 \\
optB88-vdW    & 1188 & 486 & 2.80 & 55.70 \\
revPBE-D4     & 1220 & 500 & 0.30 & 16.60 \\
revPBE-D3(0)  & 1220 & 500 & 0.30 & 18.20 \\
revPBE-D3(BJ) & 1220 & 500 & 0.50 & 17.70 \\
revPBE-DRSLL  & 1220 & 500 & 0.60 & 27.60 \\
revPBE-rVV10  & 1220 & 500 & 0.50 & 16.20 \\
\midrule
\multicolumn{5}{l}{\textit{meta-GGAs}} \\
M06L-D3(0)    & 1208 & 500 & 0.40 & 44.20 \\
r$^2$SCAN-D4  & 1194 & 496 & 1.30 & 46.30 \\
r$^2$SCAN-D3(BJ) & 1200 & 498 & 0.60 & 21.60 \\
r$^2$SCAN     & 1194 & 495 & 0.30 & 23.70 \\
B97M-rV       & 1150 & 500 & 0.30 & 15.80 \\
\midrule
\multicolumn{5}{l}{\textit{Hybrids}} \\
B3LYP-D3(0)   & 1100 & 350 & 0.60 & 40.40 \\
revPBE0-D3(0) & 1046 & 334 & 1.10 & 21.70 \\
HSE06-D3(0)   & 1122 & 250 & 0.80 & 52.10 \\
\midrule
\multicolumn{5}{l}{\textit{Beyond-DFT}} \\
RPA/TZ                        & 1009 & 497 & 0.70  & 23.20 \\
\multicolumn{5}{l}{RPA/QZ} \\
\quad Baseline (RPA/TZ)       & 1009 & 497 & 0.70  & 23.20 \\
\quad $\Delta$ (RPA/TZ$\to$RPA/QZ) & \multicolumn{2}{c}{243} & 0.20 & 10.90 \\
\bottomrule
\end{tabular}

\vspace{0.2cm}
\end{table}

\newpage
\phantom{.}

\newpage

\subsection{Computational Cost}

\begin{figure}[htp!]
    \centering
    \includegraphics[width=0.65\linewidth]{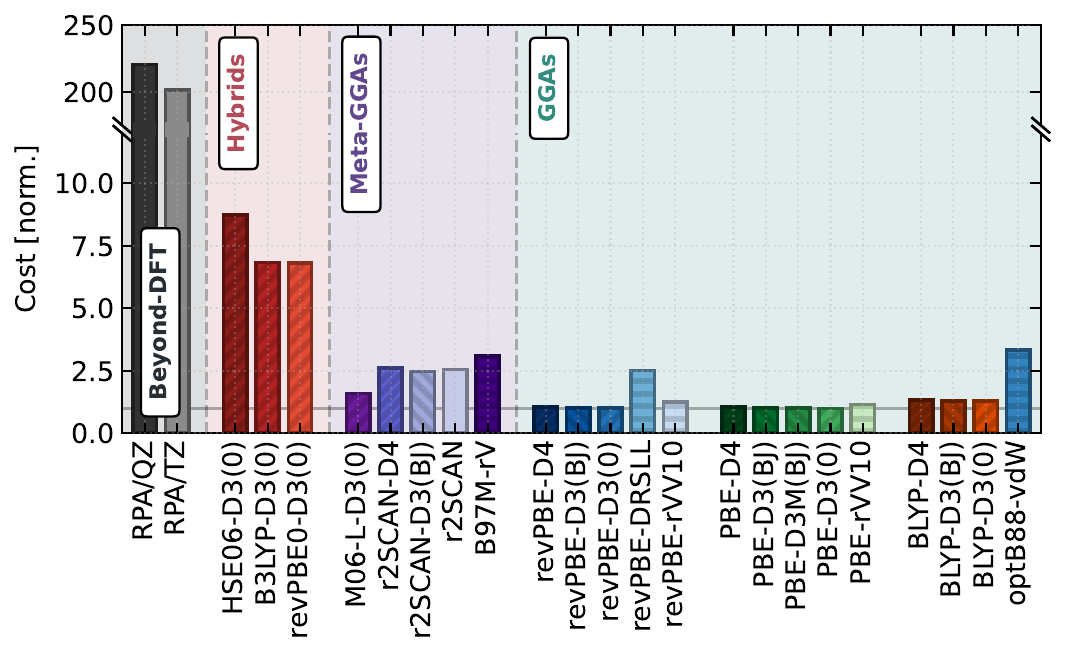}
    \caption{\textbf{Approximate computational cost of constructing the training dataset for the MLP model at each level of theory considered in this work.}
    The cost corresponds to the total compute required to generate each functional's respective dataset, normalized by the cost of the revPBE-D3(0) GGA functional.
    }
    \label{fig:si_cost_stuff}
\end{figure}

In Figure~\ref{fig:si_cost_stuff}, we report the computational cost of each level of theory considered in this work.
The costs reported here correspond to the total cost of generating the reference training datasets for the MLP models at each level of theory, rather than the cost of an individual single-point evaluation.
The different XC approximations exhibit markedly different costs, with hybrids being overall more expensive than functionals on earlier rungs of Jacob's Ladder, and RPA remaining substantially more expensive than any of the DFT functionals considered.
We note that the cost of the RPA/QZ dataset is only marginally higher than that of RPA/TZ, despite the substantially larger basis set.
This is due to the efficient $\Delta$-learning approach adopted in this work.
Rather than evaluating every structure at the QZ level, the RPA/TZ dataset is corrected using explicit QZ--TZ recalculations performed on a small subset of structures.
These numbers are intended to be illustrative, since more efficient computational frameworks exist and the settings used here could be tuned further for any given method.

\newpage

\section{Additional analyses}

\subsection{Interfacial Water}

\textbf{Density profiles with flexible graphene walls}

\begin{figure*}[htp!]
    \centering
    \includegraphics[width=0.55\linewidth]{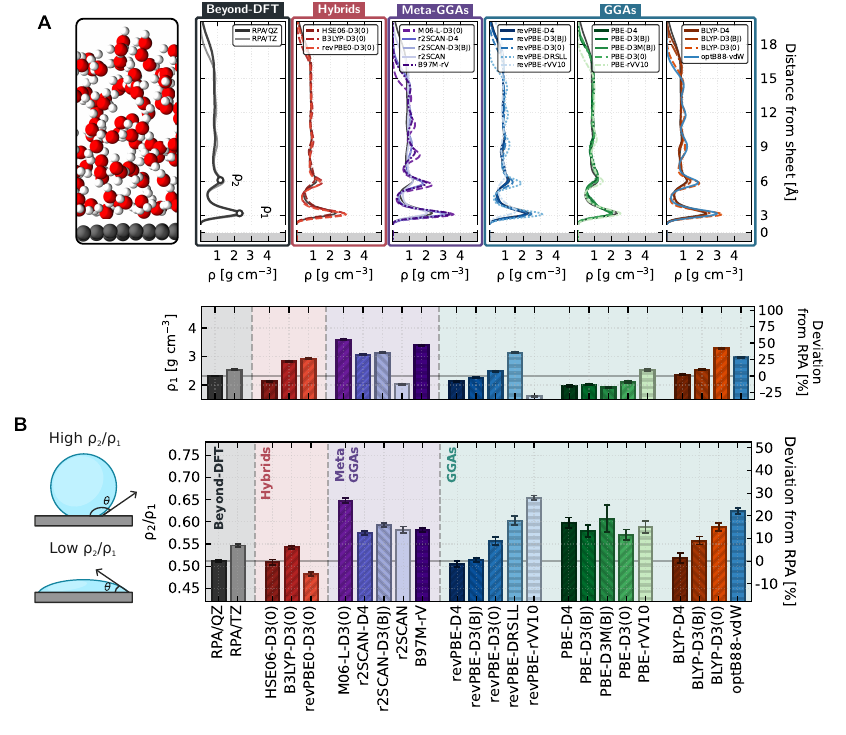}
    \caption{\textbf{Big impact of the XC approximation on interfacial water structuring and wettability at the graphene--water interface, with flexible graphene walls.}
    (A) Water oxygen density profiles as a function of the distance from the flexible graphene sheet, computed with RPA/QZ (our reference) and RPA/TZ, and grouped by XC family: hybrids, meta-GGAs, and GGAs, each paired with a variety of dispersion corrections.
    A zoomed-in view of the first interfacial peak density, $\rho_1$, is shown below for each functional, alongside its percentage deviation from the RPA reference.
    (B) Ratio of the second to first interfacial peak densities, $\rho_2/\rho_1$, used as a structural proxy for wettability, for each XC approximation.
    Percentage deviations from the RPA reference are shown alongside, with the RPA reference indicated by a horizontal dashed line where appropriate.
    }
    \label{fig:si_fig1_flexible}
\end{figure*}

In Figure~\ref{fig:si_fig1_flexible}, we replicate the analysis of Figure~1 in the main text, but for flexible graphene walls.
Density profiles are computed relative to the graphene sheet's average (equilibrium) position, as is conventionally done.
Because the graphene wall now fluctuates, an instantaneous treatment of the interface, such as the Willard-Chandler method~\cite{wc_2010}, could also be considered here.
Nonetheless, the magnitude of the variation across XC approximations is considerably larger than any difference arising from the choice of interface definition, and it is this sensitivity to the level of theory, rather than to the definition itself, that is the focus of this work.

\newpage
\textbf{Friction with rigid graphene walls}

\begin{figure}[htp!]
    \centering
    \includegraphics[width=0.65\linewidth]{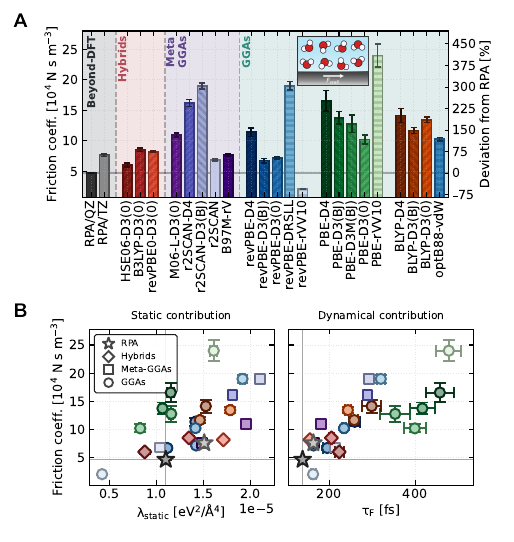}
    \caption{\textbf{Big impact of the XC approximation on the friction coefficient at the graphene--water interface, with rigid graphene walls.}
    (A) Friction coefficients computed with RPA/QZ (our reference) and RPA/TZ, and grouped by XC family: hybrids, meta-GGAs, and GGAs, each paired with a variety of dispersion corrections.
    The secondary y-axis shows the percentage deviation of each XC approximation from the RPA reference.
    (B) Decomposition of the friction coefficient into its static and dynamical contributions, shown as the friction coefficient plotted against $\lambda_{\textrm{static}}$, measure of the corrugation of the graphene--water energy landscape (left), and against $\tau_\textrm{F}$, the force correlation time capturing the relaxation of lateral forces at the interface (right).
    Colors in (B) follow the same coding as in (A), and the RPA reference is indicated by horizontal (and vertical) solid lines.
    }
    \label{fig:si_fig2_rigid}
\end{figure}

In Figure~\ref{fig:si_fig2_rigid}, we replicate the analysis of Figure~2 in the main text, but for rigid graphene walls.
As seen, the resulting friction coefficients are systematically larger than those obtained with flexible walls in the main text, consistent with previous comparisons between rigid and flexible graphene models~\cite{flt_nanotubes}.

\newpage
\textbf{Air--Water VSFG spectra}

\begin{figure*}[htp!]
    \centering
    \includegraphics[width=0.85\linewidth]{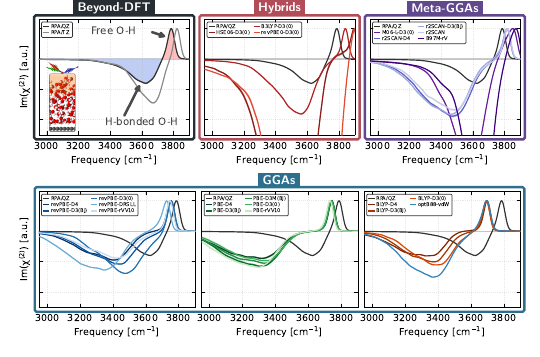}
    \caption{\textbf{Big impact of the XC approximation on the VSFG spectrum of the air--water interface.}
    Theoretical $\mathrm{Im}(\chi^{(2)})$ spectra computed with RPA/QZ (our reference) and RPA/TZ and grouped by XC family: hybrids, meta-GGAs, and GGAs, each paired with a variety of dispersion corrections, using the ssVVCF methodology~\cite{ssvcf_2015}.
    The RPA/QZ spectrum is shown in every panel for reference, and all spectra are normalized to their maximum value.
    The inset in the RPA panel shows a schematic of the system considered, highlighting the topmost interfacial water layer, whose dangling and H-bonded O-H groups dominate the spectral response.
    }
    \label{fig:si_fig3_airwater}
\end{figure*}

In Figure~\ref{fig:si_fig3_airwater}, we extend our analysis to the air--water interface, replicating the VSFG comparison of Figure~3 in the main text for this alternative, graphene-free interface.
The dangling (free) O-H frequency at the air--water interface is blueshifted relative to the graphene--water interface (Figure~3, main text), across all functionals considered, consistent with previous studies~\cite{ohto_gra_hbn_2018, wang_ange_2024}.

\newpage
\subsection{Comparison With Experiments}

\vspace{-0.5cm}

\textbf{Friction}

Experimental studies of liquid-solid friction do not report friction coefficients directly, but rather slip lengths, the two being related by $b = \eta/\lambda$ with $\eta$ the shear viscosity of the fluid.
We have not computed the viscosity for each of our models, as its convergence requires considerably longer sampling than the quantities considered here~\cite{viscosity_abascal_2010}, so we use the experimental value for water at $300$~K ($\eta = 0.855 \times 10^{-3}$~Pa~s), which makes the comparison indicative rather than quantitative.
Our RPA reference, $\lambda \approx 3.4 \times 10^{4}$~N~s~m$^{-3}$, then corresponds to a slip length of roughly $25$~nm, well within the range reported for water on flat graphitic surfaces.
Capillary filling in graphene nanochannels gives a median slip length of $16$~nm~\cite{Xie2018_natnano}, a three-dimensional analysis of the same geometry gives $30$-$40$~nm~\cite{chen_slip_2022}, and measurements in graphite capillaries give values towards the upper end of this range~\cite{boya_2021_natcomms_friction}.
We would caution against reading too much into this agreement, given that atomic force microscopy on multilayer graphite gives instead $4$-$8$~nm~\cite{maali_2008, li_ming_2022}.
Reported values therefore scatter by more than an order of magnitude, a range comparable to the one our own functionals span, so that a measured slip length cannot at present be mapped onto a particular level of theory.
Much of that spread appears to be extrinsic rather than intrinsic to the interface, since colloidal-probe measurements on the same material give a slip length of about $1$~nm for graphene on SiO$_2$ but about $13$~nm for graphene on silanized silica~\cite{li_ming_2022}, and the large variation reported for graphene-covered silica grooves has been attributed to differences in graphene quality and in the graphene-substrate interaction~\cite{Xie2018_natnano}.
A further caution is that electronic contributions to friction, absent from any classical description, have been proposed as a significant term for water on graphite, with an estimated coefficient of comparable magnitude to the values discussed here~\cite{kavokine_friction_2022, sutter2026_rev}.

This relative agreement is not found in carbon nanotubes, where measured flow rates~\cite{majumder_2005, holt_2006, secchi_2016_nature} imply friction coefficients up to two orders of magnitude below those obtained from simulation~\cite{flt_nanotubes}.
Part of that discrepancy lies in the measurements themselves, which disagree substantially with one another, a spread attributed to the difficulty of interpreting membrane-based transport assays and to the strong radius dependence of the slip length, which ranges from $30$~nm in the largest tubes to $300$~nm in the smallest~\cite{secchi_2016_nature, sutter2026_rev}.
A further contribution is specific to the tube geometry, since measurements on individual tubes are typically made on multiwalled samples~\cite{secchi_2016_nature}, where interlayer coupling allows the same electronic contributions to arise~\cite{kavokine_friction_2022, sutter2026_rev}.

\textbf{Sum-Frequency Generation Spectroscopy}

\begin{figure*}[htp!]
    \centering
    \includegraphics[width=0.70\linewidth]{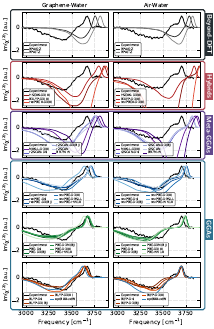}
    \caption{\textbf{Computed VSFG spectra compared with heterodyne-detected VSFG measurements.}
    Theoretical $\mathrm{Im}(\chi^{(2)})$ spectra at the graphene--water (left) and air--water (right) interfaces, grouped by XC family and computed using the ssVVCF methodology~\cite{ssvcf_2015}.
    The measured spectra, shown in black in every panel, are taken from the data reported in Ref.~\citenum{wang_ange_2024} for a freely suspended graphene--water interface.
    All spectra are normalized to their dangling O-H maximum, and shaded regions show the standard error across independent runs.
    }
    \label{fig:si_sfg_expt}
\end{figure*}

The suspended graphene--water interface has been characterized by heterodyne-detected VSFG spectroscopy~\cite{wang_ange_2024}, which makes it one of the few solid--liquid systems where computed $\mathrm{Im}(\chi^{(2)})$ spectra can be set against measurement for the same interface.
Two such measurements have been reported and disagree on the redshift of the dangling O-H peak, among other features~\cite{xu_tian_2023, wang_ange_2024}.
We compare against the more recent one~\cite{wang_ange_2024}, which finds no appreciable bulk contribution or graphene VSFG response and concludes that the origins of these contributions in the earlier study should be reevaluated.
In Figure~\ref{fig:si_sfg_expt} we overlay the measured spectra at both the graphene--water and air--water interfaces on our computed spectra, grouped by functional family, with the experimental curves obtained from Ref.~\citenum{wang_ange_2024} and normalized to the dangling O-H maximum.

Two systematic differences are apparent, neither of which should be read as a failure of a particular XC approximation.
The computed dangling O-H frequencies span more than $250$~cm$^{-1}$, from $3620$~cm$^{-1}$ for optB88-vdW to $3876$~cm$^{-1}$ for revPBE0-D3(0), bracketing the measured value of $3670$~cm$^{-1}$, while the computed hydrogen-bonded band comes out consistently narrower than the measured one.
We note that our simulations treat the nuclei classically, and therefore omit the nuclear quantum effects that substantially lower the frequency of a mode as anharmonic as the O-H stretch.
The computed frequencies are consequently expected to be systematically blueshifted, an offset also present in the AIMD and machine-learned simulations reported alongside the measurements in Ref.~\citenum{wang_ange_2024}.
The narrower band may have a separate origin, since the ssVVCF algorithm we use neglects the intermolecular and intramolecular couplings of the O-H stretch, which contribute to the breadth of the hydrogen-bonded feature.
A further difference worth noting is that the measured sheet is freely suspended and able to fluctuate, whereas we hold ours rigid to isolate the effect of the XC approximation on the interfacial water structure.

Absolute frequencies are therefore not the right basis for comparison.
The quantity we use in the main text is instead the redshift of the dangling O-H peak between the two interfaces, given that both are computed with the same method, and in our case from the same simulation cell, so that the systematic offset largely cancels and what remains reflects the graphene--water interaction itself.
The values obtained for every functional are collected in Table~\ref{tab:sfg_summary}.

\newpage

\newpage

\begin{table}[htp!]
\centering
\caption{Dangling and hydrogen-bonded O-H features of the computed VSFG spectra at the graphene--water and air--water interfaces.
$\omega_{\mathrm{dang}}$ is the position of the dangling O--H maximum, while $\mathrm{FWHM}{_\mathrm{HB}}$ is the width of the hydrogen-bonded band, defined as its full width at half maximum.
$\Delta\omega_{\mathrm{dang}} = \omega_{\mathrm{dang}}^{\mathrm{air}} - \omega_{\mathrm{dang}}^{\mathrm{gra}}$ is the redshift of the dangling O--H peak upon going from the air--water to the graphene--water interface.
Experimental values are extracted from the spectra of Ref.~\citenum{wang_ange_2024}, and differ slightly from the $30 \pm 10$~cm$^{-1}$ reported there due to the analysis used.}

\label{tab:sfg_summary}
\vspace{0.4cm}
\begin{tabular}{lccccc}
\toprule
 & \multicolumn{2}{c}{Graphene--Water} & \multicolumn{2}{c}{Air--Water} & \\
\cmidrule(lr){2-3}\cmidrule(lr){4-5}
 & $\omega_{\mathrm{dang}}$ & FWHM$_{\mathrm{HB}}$ & $\omega_{\mathrm{dang}}$ & FWHM$_{\mathrm{HB}}$ & \\
\multirow{-3}{*}{Model} & [cm$^{-1}$] & [cm$^{-1}$] & [cm$^{-1}$] & [cm$^{-1}$] & \multirow{-3}{*}{\shortstack{$\Delta\omega_{\mathrm{dang}}$\\{}[cm$^{-1}$]}} \\
\midrule
\multicolumn{6}{l}{\textit{GGAs}} \\
PBE-D4                 & 3670 & 426 & 3743 & 416 & 72 \\
PBE-D3(0)              & 3682 & 424 & 3737 & 393 & 56 \\
PBE-D3(BJ)             & 3670 & 452 & 3743 & 400 & 72 \\
PBE-D3M(BJ)            & 3676 & 462 & 3737 & 404 & 61 \\
PBE-rVV10              & 3637 & 441 & 3732 & 400 & 94 \\
BLYP-D4                & 3637 & 354 & 3698 & 350 & 61 \\
BLYP-D3(0)             & 3648 & 331 & 3693 & 302 & 44 \\
BLYP-D3(BJ)            & 3643 & 361 & 3693 & 317 & 50 \\
optB88-vdW             & 3620 & 354 & 3693 & 330 & 72 \\
revPBE-D4              & 3704 & 382 & 3748 & 346 & 44 \\
revPBE-D3(0)           & 3715 & 340 & 3754 & 295 & 39 \\
revPBE-D3(BJ)          & 3709 & 383 & 3754 & 346 & 44 \\
revPBE-DRSLL           & 3637 & 411 & 3726 & 378 & 89 \\
revPBE-rVV10           & 3704 & 364 & 3748 & 338 & 44 \\
\midrule
\multicolumn{6}{l}{\textit{meta-GGAs}} \\
M06L-D3(0)             & 3770 & 247 & 3848 & 233 & 78 \\
r$^2$SCAN-D4           & 3754 & 384 & 3815 & 358 & 61 \\
r$^2$SCAN-D3(BJ)       & 3754 & 391 & 3815 & 366 & 61 \\
r$^2$SCAN              & 3770 & 373 & 3821 & 353 & 50 \\
B97M-rV                & 3810 & 191 & 3871 & 191 & 61 \\
\midrule
\multicolumn{6}{l}{\textit{Hybrids}} \\
B3LYP-D3(0)            & 3798 & 277 & 3837 & 270 & 39 \\
revPBE0-D3(0)          & 3876 & 268 & 3898 & 262 & 22 \\
HSE06-D3(0)            & 3854 & 350 & 3893 & 347 & 39 \\
\midrule
\multicolumn{6}{l}{\textit{Beyond-DFT}} \\
RPA/TZ                 & 3815 & 212 & 3821 & 188 & 6 \\
RPA/QZ                 & 3754 & 193 & 3787 & 190 & 33 \\
\midrule
\multicolumn{6}{l}{\textit{Experiment}} \\
Experiment (Ref.~\citenum{wang_ange_2024})            & 3671 & 290 & 3699 & 339 & 28 \\
\bottomrule
\end{tabular}
\vspace{0.2cm}
\end{table}

\newpage

\subsection{Individual and Overall Scores for All XC Approximations}

While Figure~4 in the main text shows the overall score for a representative subset of XC approximations, here we provide the full breakdown of individual and overall scores for all functionals considered in this work.

\begin{figure*}[htp!]
    \centering
    \includegraphics[width=\linewidth]{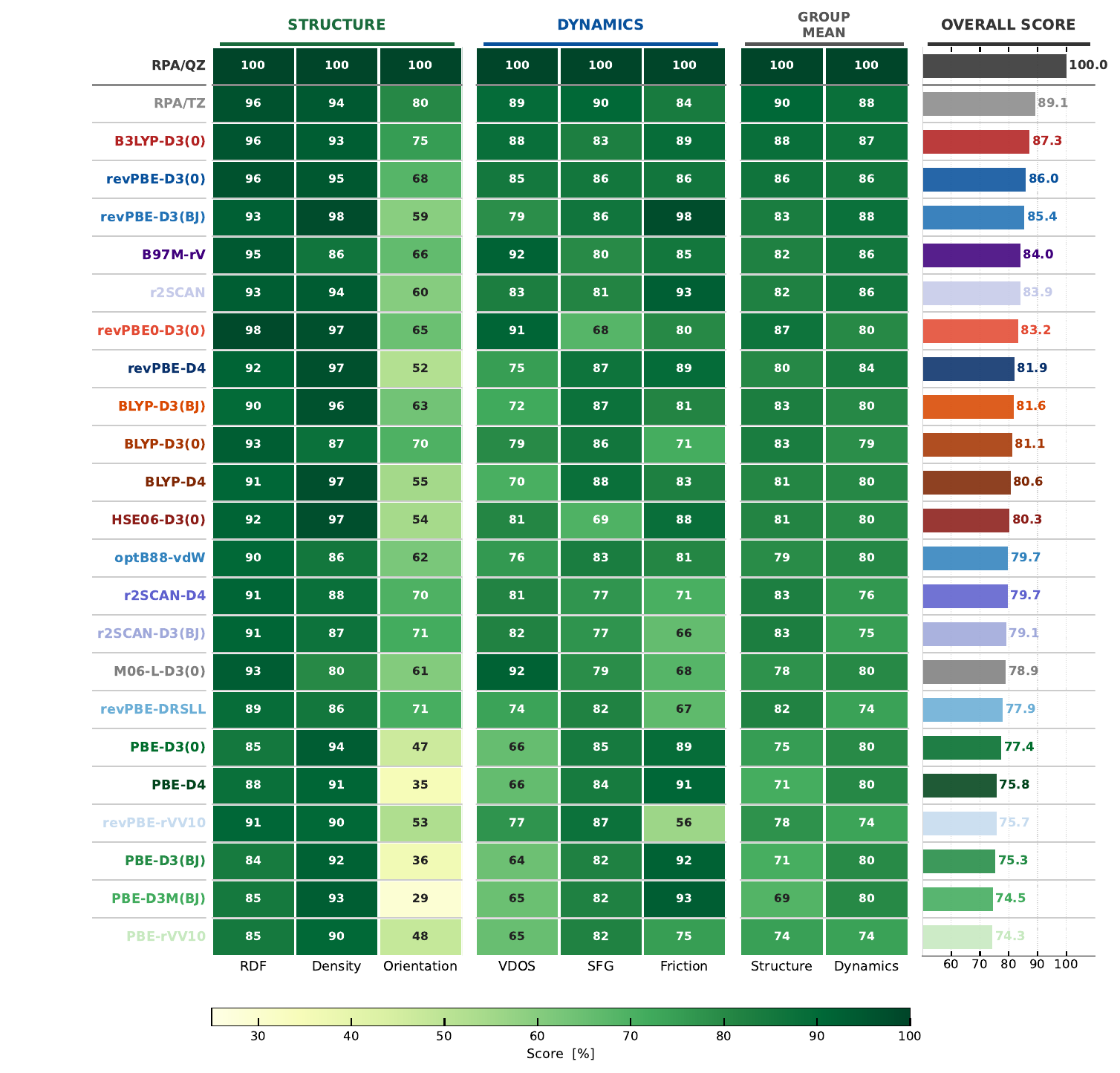}
    \caption{\textbf{Full breakdown of individual and overall scores for all XC approximations.}
    Heatmap of individual scores for each structural observable (RDF, density, orientation) and each dynamical observable (VDOS, VSFG, friction), together with the group mean for each category, for every XC approximation considered in this work.
    Rows are sorted by overall score, shown at right and computed as the mean across all six individual scores.
    Color shading reflects the score value, from 25\% (light) to 100\% (dark), as indicated in the colorbar.
    }
    \label{fig:si_fig4}
\end{figure*}
\newpage

\subsection*{Static Adsorption Energies versus Finite-Temperature Observables}

Previous assessments of XC functionals at solid--water interfaces have relied largely on static quantities, most commonly the adsorption energy of a single water molecule on the surface.
Here we test whether such a benchmark is able to anticipate the finite-temperature behavior considered in the main text.
To this end, we computed the adsorption energy of a single water molecule on graphene, $E_{\textrm{ads}}^{\textrm{min}}$, in the 0-, 1-, and 2-leg configurations, which correspond to zero, one, and two O-H bonds pointing toward the sheet, respectively (see Figure~\ref{fig:si_rpa_binding_energy}).
We note that these were obtained directly at each level of electronic structure theory, rather than with the corresponding MLPs, using the same settings as for the training data (see Section~\ref{sec:elect_settings}).
In each case, $E_{\textrm{ads}}^{\textrm{min}}$ is taken as the minimum of the corresponding binding curve.

\begin{figure*}[htp!]
    \centering
    \includegraphics[width=0.9\linewidth]{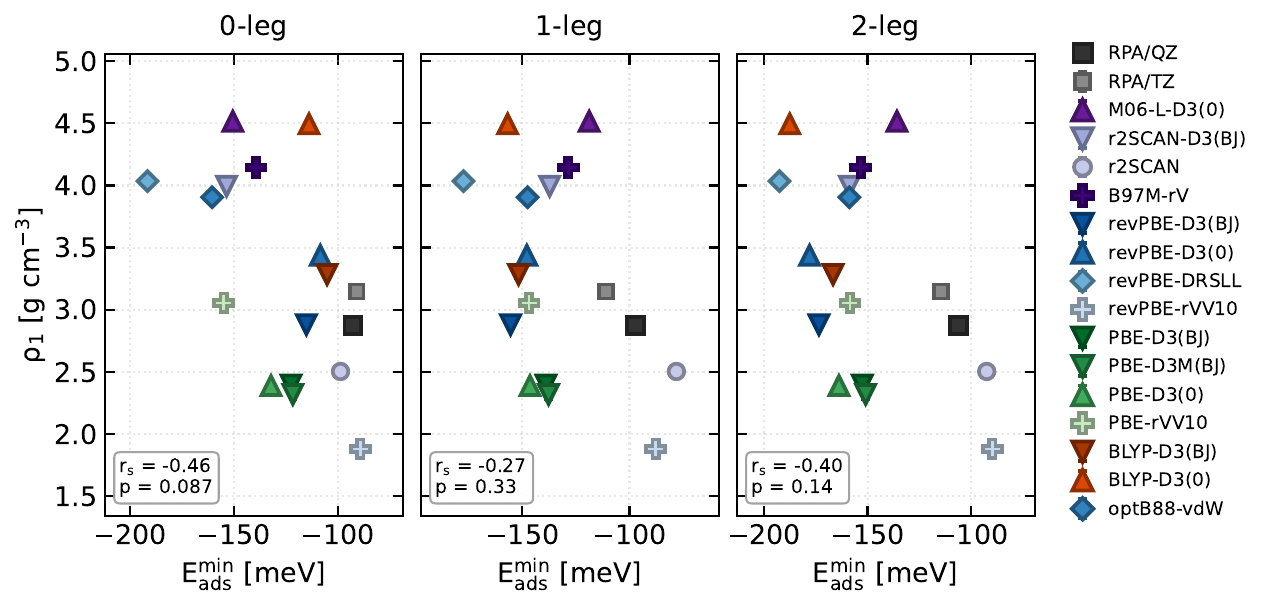}
    \caption{\textbf{Static adsorption energies compared with the finite-temperature contact-layer density.}
    Contact-layer density $\rho_1$ plotted against the adsorption energy of a single water molecule on graphene, $E_{\textrm{ads}}^{\textrm{min}}$, for the 0-leg (left), 1-leg (middle), and 2-leg (right) configurations.
    Each panel reports the Spearman rank correlation coefficient $r_s$ and its $p$-value, computed over the DFT functionals only.
    }
    \label{fig:si_eads_rho1}
\end{figure*}

\begin{figure*}[htp!]
    \centering
    \includegraphics[width=0.9\linewidth]{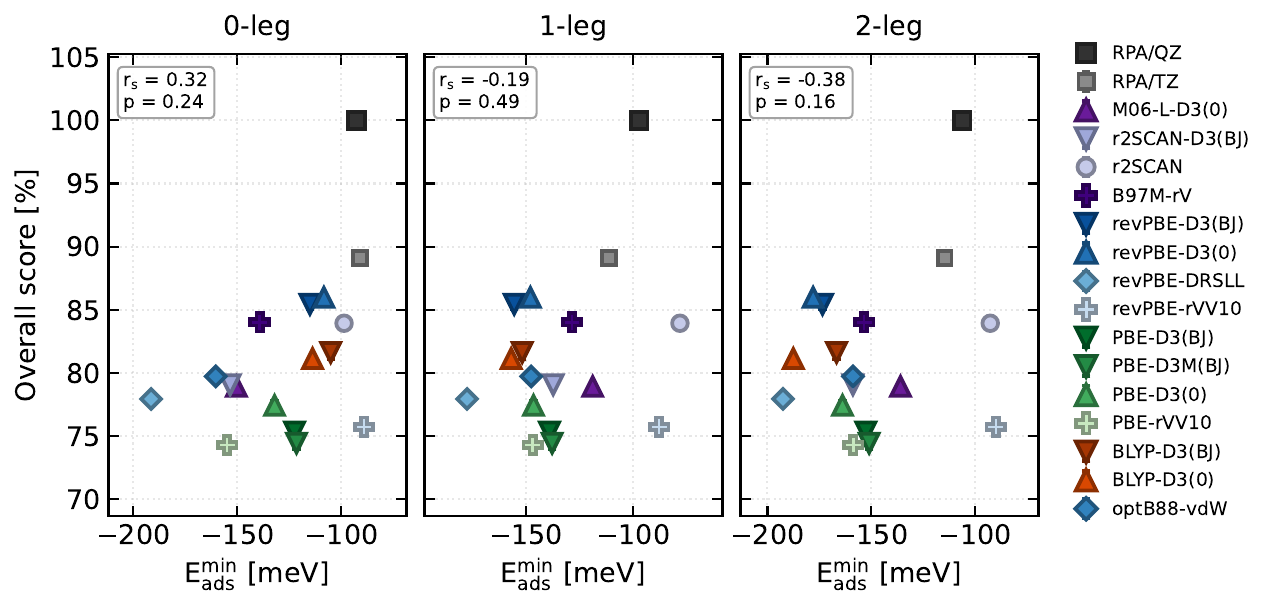}
    \caption{\textbf{Static adsorption energies compared with the overall score.}
    Overall score from the multi-observable scoring framework plotted against the adsorption energy of a single water molecule on graphene, $E_{\textrm{ads}}^{\textrm{min}}$, for the 0-leg (left), 1-leg (middle), and 2-leg (right) configurations.
    Each panel reports the Spearman rank correlation coefficient $r_s$ and its $p$-value, computed over the DFT functionals only.
    }
    \label{fig:si_eads_score}
\end{figure*}

As shown in Figure~\ref{fig:si_eads_rho1}, the RPA/QZ reference lies at the weak-binding end of the distribution, in line with diffusion Monte Carlo~\cite{brandenburg_2019}, whereas almost every DFT functional binds the water molecule more strongly.
We find that these adsorption energies correlate only moderately with the contact-layer density $\rho_1$ obtained from the finite-temperature simulations, with the correlation reaching significance in two of the three configurations.
Notably, however, the overall score shows no significant correlation with the adsorption energy in any of the three configurations (Figure~\ref{fig:si_eads_score}).
In other words, even where the adsorption energy tracks a structural property of the interface, it is not a reliable guide to how faithfully a functional describes the interface as a whole at finite temperature.

These data also bear on the origin of the elevated $\rho_2/\rho_1$ ratios discussed in the main text.
Most functionals predict a more hydrophobic interface than RPA, despite binding a single water molecule to graphene more strongly.
This suggests that water--water interactions are overestimated by even more, although we emphasize that a single-molecule minimum-energy structure is not equivalent to the adhesion of a liquid contact layer, and so this inference should be read with some caution.

\newpage

\newpage
\section{Data Analysis} 
\subsection{Friction Coefficient}
The friction coefficient $\lambda$ is defined as the proportionality constant relating the total lateral force exerted by the solid (here, the graphene surface) on the water layer parallel to it, $\mathbf{F}_{\textrm{wall}}$, normalized by the surface area $A$, to the fluid velocity at the interface, $\mathbf{v}_{\textrm{wall}}$:
\begin{equation}
    \lambda \mathbf{v}_{\textrm{wall}} = \frac{\mathbf{F}_{\textrm{wall}}}{A}.
\end{equation}
From linear response theory, $\lambda$ can be computed directly from equilibrium MD  simulations via the Green-Kubo relation~\cite{bocquet_barrat_friction_1994, bocquet_barrat_2014},
\begin{equation}
    \lambda_{\textrm{GK}} (t) = \frac{1}{A k_{\textrm{B}}T} \int_{0}^{t} \langle \mathbf{F}_{\textrm{wall}} (t^{\prime}) \cdot  \mathbf{F}_{\textrm{wall}} (0)  \rangle dt^{\prime},
\end{equation}
where $k_{\textrm{B}}$ is the Boltzmann constant, $T$ is the temperature, and the brackets denote an ensemble average.
$\mathbf{F}_{\textrm{wall}}$ is computed as the summed lateral force of all solid atoms on the water layer, sampled every 1~fs, with both in-plane components $(x,y)$ included.
The friction coefficient is formally recovered as $\lambda = \lim_{t \to \infty} \lambda_{\textrm{GK}} (t)$.
In practice, however, Green-Kubo integrals vanish at long times in finite systems~\cite{friction_fs_1997} (the so-called ``plateau problem''~\cite{plateau_espanol_2019}) complicating their direct evaluation.
We circumvent this by fitting the analytical expression~\cite{joly_analytical_2019}
\begin{equation}
    \Lambda (t) = \lambda_0 \left( e^{-\frac{t}{t_1}} - e^{-\frac{t}{t_2}}  \right)
\end{equation}
to the computed integrals, from which the finite-size corrected friction is obtained as
\begin{equation}
\lambda = \lambda_0 \, \frac{1 - u}{1 + u}, \qquad u = t_2/t_1. 
\end{equation}

Following the approach proposed in Ref.~\citenum{tocci_static_dyn_2020}, the friction coefficient can be further decomposed into a static and a dynamic contribution by rewriting the Green-Kubo relation as
\begin{equation}
    \lambda = \frac{1}{A k_{\textrm{B}} T} \langle \mathbf{F}_{\textrm{wall}}^2 \rangle \times \tau_{\textrm{F}},
\end{equation}
where $\langle \mathbf{F}_{\textrm{wall}}^2 \rangle$ is the mean squared lateral force acting on the fluid, obtained as a simple time average over the trajectory, and $\tau_{\textrm{F}}$ is the force correlation time, defined as
\begin{equation}
    \tau_{\textrm{F}} = \int_{0}^{\infty} \frac{\langle \mathbf{F}_{\textrm{wall}} (0) \cdot \mathbf{F}_{\textrm{wall}} (t) \rangle}{\langle \mathbf{F}_{\textrm{wall}}^2 \rangle} \, dt.
\end{equation}
This decomposition separates $\lambda$ into a static contribution, $\lambda_{\textrm{static}} = \langle \mathbf{F}_{\textrm{wall}}^2 \rangle / A$, reflecting the corrugation of the graphene--water energy landscape, and a dynamic contribution, $\tau_{\textrm{F}}$, reflecting the timescale over which lateral forces at the interface decorrelate.

\subsection{Vibrational Sum-Frequency Generation Spectroscopy}

The resonant VSFG spectra were computed via the surface-specific velocity–velocity correlation function (ssVVCF) approach~\cite{ssvcf_2015}.
Within this framework, the $xxz$ component of the second-order susceptibility, which corresponds to the SSP polarization combination most commonly measured in experiment, can be given by

\begin{equation}
\chi^{(2)}_{xxz}(\omega) = \frac{Q(\omega)\mu'_\text{str}\alpha'_\text{str}}{i\omega^2} \int_0^{\infty} dt \, e^{-i\omega t} \left\langle \sum_{i} g_\text{ds}(z_i(0)) \, \dot{\mathbf{r}}^{\text{OH}}_{z,i}(0) \, \frac{\dot{\mathbf{r}}^{\text{OH}}_{i}(t) \cdot \mathbf{r}^{\text{OH}}_{i}(t)}{|\mathbf{r}^{\,\text{OH}}_i(t)|} \right\rangle,
\end{equation}

where $\mu'_\text{str}$ and $\alpha'_\text{str}$ are the dipole moment and polarizability derivatives of the O--H stretching mode, respectively, and $\mathbf{r}^{\text{OH}}_{i}$ and $\dot{\mathbf{r}}^{\text{OH}}_{i}$ denote the unit vector and velocity along the $i$th O--H bond, and $Q(\omega) = \beta \hbar\omega/(1-e^{-\beta\hbar\omega})$ is the quantum correction factor~\cite{berens_quantum_corr_fact_1981}, where $\beta=1/k_{\textrm{B}}T$.
The truncation function $g_\text{ds}(z_i)$ selects interfacial water molecules and is defined as unity for $z_i < z_\text{ds}$ and zero otherwise, where $z_\text{ds}$ is the position of the dividing surface.
We specifically focused on the topmost interfacial water layer, which dominates the spectral response as shown in our previous work~\cite{ions_gra_advin_2025}.
Computing VSFG spectra is considerably more demanding than bulk spectroscopies.
The signal arises from only a few molecular layers at the interface, and extensive sampling is required to resolve the subtle spectral features that survive cancellation of bulk contributions~\cite{ssvcf_2015, morita_book, yair_sfg_2023}.
The use of MLPs was therefore critical here, as it enabled multi-nanosecond trajectories that are well beyond the reach of direct ab initio MD and necessary to converge the spectral differences central to our analysis.

\newpage

\section{Multi-Observable Scoring Framework} \label{sec:scoring}
Defining a scoring metric provides a practical and accessible means of ranking models trained on different DFT functionals.
To this end, we establish a validation protocol tailored to solid--liquid interfacial environments, assessing model performance across six categories of observables: radial distribution functions (RDFs), vibrational density of states (VDOS), density and orientation profiles of the liquid at the interface, vibrational sum-frequency generation (VSFG) spectroscopy, and the solid--liquid friction coefficient.
The overall score for a given functional is obtained by averaging its individual scores across all six categories, each compared directly against our RPA/QZ reference.
All scores, whether comparing functions or scalars, use the same symmetric normalization, in which the difference between reference and test is divided by the sum of their magnitudes rather than by the reference alone.
This treats over- and under-estimation on an equal footing, so that a quantity predicted twice too large scores the same as one predicted twice too small, which matters for quantities such as the friction coefficient that vary by factors rather than by percentages across functionals.
It also avoids the saturation of a reference-normalized relative error, which assigns a score of zero to any deviation beyond 100\% and so cannot distinguish a functional that is twice off from one that is five times off.
Together, these properties let us assess structural, dynamical, and spectroscopic behavior across all functionals in solid--liquid environments.
Each individual contribution is described in detail below.

\vspace{-0.5cm}

\subsection*{Radial Distribution Functions}

RDFs provide detailed insight into the pairwise structural arrangement of atoms within the system.
To score the different DFT functionals, we compute all RDFs among the liquid atoms (here O and H).
The solid--liquid RDF is not included here, as it is already captured by the density profile metric described below.
We evaluate these RDFs in two distinct environments, bulk water and interfacial water, since accurately describing both is critical for modeling solid--liquid interfaces.
For a system with $N$ liquid atom species, this yields $2\cdot\binom{N+1}{2}$ RDFs, which can be directly compared between the RPA/QZ reference and each XC approximation using a suitable norm $d^{\textrm{RDF}}$,
\begin{equation}
    d^{\textrm{RDF}} = 1-\frac{\int_{0}^{+\infty} | g^{\textrm{RPA}} (r) - g^{\textrm{XC}} (r)|dr   }{ \int_{0}^{+\infty} | g^{\textrm{RPA}} (r)| dr +  \int_{0}^{+\infty} | g^{\textrm{XC}} (r)| dr }
\end{equation}
which measures the similarity between two RDFs, ranging from 0 (maximally different) to 1 (identical).
For each environment, we average $d^{\textrm{RDF}}$ over the $\binom{N+1}{2}$ species pairs to obtain a single combined score for bulk water and one for interfacial water.
Averaging these two environment scores then yields one overall value per functional.
Converting this value to a percentage provides a condensed measure of RDF performance relative to the RPA/QZ reference.

\subsection*{Vibrational Density of States}

The VDOS provides detailed insight into the vibrational and dynamical properties of the system, capturing how atomic motions are distributed across frequencies.
It is obtained via the Fourier transform of the velocity autocorrelation function.
The VDOS can be computed separately for each component of the system, probing distinct vibrational processes across a broad frequency range.
For a system with $N$ species, this yields $N$ species-resolved VDOS (here we consider O and H), computed for both the reference and each XC approximation, and evaluated in the same two environments as the RDFs, bulk water and interfacial water.
Using a norm $d^{\textrm{VDOS}}$ analogous to that used for the RDFs,

\begin{equation}
    d^{\textrm{VDOS}} = 1-\frac{\int_{0}^{+\infty} | f^{\textrm{RPA}} (\omega) - f^{\textrm{XC}} (\omega)|d\omega   }{ \int_{0}^{+\infty} | f^{\textrm{RPA}} (\omega)| d\omega +  \int_{0}^{+\infty} | f^{\textrm{XC}} (\omega)| d\omega }
\end{equation}

we obtain a measure of similarity between the RPA/QZ and XC VDOS spectra for each species and environment.
Averaging over species and then over environments, and converting the result to a percentage, yields the overall score for a given XC approximation.

\subsection*{Density profiles}

The density profiles of a liquid in contact with a solid provide detailed insight into the solid--liquid structuring.
For a system with $N$ species, this yields $N$ species-resolved density profiles (here we consider O and H), computed for both the reference and each XC approximation.
This metric, along with the subsequent ones presented, is only computed for interfacial environments, as such a property does not apply to bulk systems.
Using a norm $d^{\textrm{DP}}$ analogous to that used for the RDFs,

\begin{equation}
    d^{\textrm{DP}} = 1-\frac{\int_{0}^{+\infty} | {\rho}^{\textrm{RPA}} (z) - {\rho}^{\textrm{XC}} (z)|dz   }{ \int_{0}^{+\infty} | {\rho}^{\textrm{RPA}} (z)| dz +  \int_{0}^{+\infty} | {\rho}^{\textrm{XC}} (z)| dz }
\end{equation}

we obtain a measure of similarity between the RPA/QZ and XC density profiles for each species at the interface.
We are additionally interested in the ratio between the first and second density peaks of the oxygen density profile ($\rho_2/\rho_1$), given its correlation with the macroscopic contact angle and wetting properties~\cite{Agosta2026}, and so score this quantity separately.
Since $\rho_2/\rho_1$ is a scalar rather than a function of $z$, we use a scalar analogue of the norm above, $d^{\rho_2/\rho_1}$, 

\begin{equation}
    d^{\rho_2/\rho_1} = 1-\frac{|(\rho_2/\rho_1)^{\textrm{RPA}} - (\rho_2/\rho_1)^{\textrm{XC}}|}{|(\rho_2/\rho_1)^{\textrm{RPA}}|+|(\rho_2/\rho_1)^{\textrm{XC}}|}
\end{equation}

which provides a measure of similarity between the RPA/QZ and XC $\rho_2/\rho_1$ ratios.
We first average $d^{\textrm{DP}}$ over the O and H density profiles to obtain a single density-profile score, then average this with $ d^{\rho_2/\rho_1}$.
Converting the result to a percentage yields the overall score for a given XC approximation.

\subsection*{Orientation profiles}

The orientational profiles capture the preferred alignment of water molecules as a function of distance from the solid surface, providing insight into interfacial ordering that complements the density profiles.
As with the density profiles, this metric is only computed for interfacial environments.
Because regions far from the surface (especially, at the air--water interface) are sparsely sampled, we weight each bin's contribution to the score by its local population, so that noisy, low-statistics regions do not dominate the comparison.
For the same reason, the comparison is further restricted to a slab of width $z_{\textrm{max}} = 11\,\textrm{\AA}$ above the graphene sheet, which corresponds to roughly three layers of water at the RPA/QZ reference level.
Beyond this distance, the orientational signal is dominated by the air--water interface and by the statistical noise associated with the small number of molecules sampled there, neither of which is informative about the solid--liquid interface we set out to describe.
Using a population-weighted norm $d^{\textrm{OP}}$ analogous to that used above,

\begin{equation}
d^{\textrm{OP}} = 1-\frac{\int_{0}^{z_{\textrm{max}}} w(z) \left| f^{\textrm{RPA}} (z) - f^{\textrm{XC}} (z)\right|dz}{ \int_{0}^{z_{\textrm{max}}} w(z) \left(\left| f^{\textrm{RPA}} (z)\right| + \left| f^{\textrm{XC}} (z)\right|\right) dz }
\end{equation}
we obtain a measure of similarity between the RPA/QZ and XC orientational profiles.
Here $f(z)$ is the orientational profile, $z$ is measured from the graphene sheet, and $w(z)$ is the local water population, taken from the RPA/QZ reference, which suppresses the contribution of sparsely sampled regions far from the surface.

\subsection*{Vibrational Sum-Frequency Generation Spectroscopy}

VSFG spectroscopy probes the vibrational response of interfacial water molecules and is highly sensitive to their structural and orientational arrangement at the surface, making it a stringent test of interfacial structure.
We compute VSFG spectra for two distinct interfaces: graphene--water and air--water.
The graphene--water spectrum is the primary target of interest, while the air--water spectrum comes essentially for free, since the slab geometry used for the graphene--water simulations exposes a water-vacuum interface on the opposite side, with a bulk-like water region in between~\cite{advincula_inter_nano_2026}.
Rather than comparing spectra point-by-point, we score each spectrum on four physically meaningful descriptors, chosen to reflect the features typically used to interpret VSFG spectra, namely the positions of the hydrogen-bonded and dangling O-H peaks ($\omega_{\mathrm{HB}}$ and $\omega_{\mathrm{dang}}$), the ratio of their intensities ($I_{\mathrm{dang}}/I_{\mathrm{HB}}$), and the width of the hydrogen-bonded band ($\mathrm{FWHM}_{\mathrm{HB}}$).
The H-bond peak is identified as the minimum of the spectrum within the analysis window, and the dangling-OH peak as the maximum.
Their respective positions, $\omega_{\textrm{HB}}$ and $\omega_{\textrm{dang}}^{\textrm{OH}}$, are compared between the reference and each XC approximation.
The intensity ratio is defined as $I_{\textrm{dang}}^{\textrm{OH}}/I_{\textrm{HB}}$, the ratio of the dangling-OH peak amplitude to the H-bond peak amplitude, and captures the relative balance between ordered and disordered interfacial water populations.
The H-bond band width is quantified as the full width at half depth of the H-bond peak, $\textrm{FWHM}_{\textrm{HB}}$.
For each descriptor $q$, we define an agreement score
\begin{equation}
    a_q = 1-\frac{|q^{\textrm{XC}}-q^{\textrm{RPA}}|}{|q^{\textrm{RPA}}|+|q^{\textrm{XC}}|}
\end{equation}
using the same normalization as the observables above, with $a_q=1$ indicating a perfect match and $a_q=0$ the least favorable outcome the metric can register.
The peak-position agreement is taken as the mean of the hydrogen-bonded and dangling-OH position agreements, and the overall SFG score is the mean of the position, intensity-ratio, and width agreements,
\begin{equation}
    d^{\textrm{SFG}} = \frac{1}{3}\left(\frac{a_{\omega_{\textrm{HB}}}+a_{\omega_{\textrm{dang}}^{\textrm{OH}}}}{2} + a_{I_{\textrm{dang}}^{\textrm{OH}}/I_{\textrm{HB}}} + a_{\textrm{FWHM}_{\textrm{HB}}}\right).
\end{equation}
For each interface, the score is first averaged over the available simulation runs.
The resulting graphene--water and air--water scores are then averaged, together with a conversion to a percentage, to yield the overall SFG score for a given XC approximation.

\begin{figure*}[htp!]
    \centering
    \includegraphics[width=0.6\linewidth]{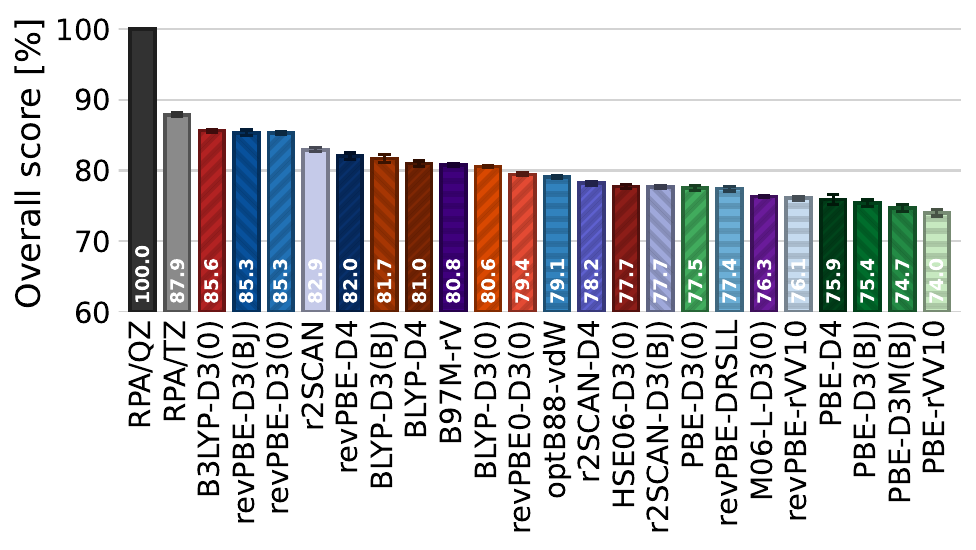}
    \caption{\textbf{Overall scores obtained with an integrated band width in place of $\textrm{FWHM}_{\textrm{HB}}$.}
    Overall score for each XC approximation, obtained by averaging over all individual observable scores, with the VSFG band width quantified as the integral of the hydrogen-bonded lobe rather than its full width at half depth.
    }
    \label{fig:si_ranking_integral}
\end{figure*}

As an additional check on the sensitivity of the VSFG score to the way the band width is defined, we rescored the spectra using the integral of the hydrogen-bonded lobe, taken over the negative region of $\textrm{Im}(\chi^{(2)})$ up to the zero crossing preceding the dangling O-H peak.
The resulting overall scores, shown in Figure~\ref{fig:si_ranking_integral}, agree with those obtained using $\textrm{FWHM}_{\textrm{HB}}$ to within 4\% for every XC approximation considered, with a Spearman rank correlation of 0.97 between the two rankings, so our conclusions do not hinge on the particular definition of the band width.

The two definitions have complementary weaknesses.
$\textrm{FWHM}_{\textrm{HB}}$ is measured out to where the band falls to half its depth, and can therefore be cut short if the shoulder near 3200-3400~cm$^{-1}$ itself reaches that half-depth threshold before the true edge of the band does.
The integral, by contrast, scales as depth $\times$ width, and so re-encodes the band depth that the intensity ratio already carries.
We take $\textrm{FWHM}_{\textrm{HB}}$ as our primary measure, since keeping the three descriptors independent matters more for a combined score.

\newpage

\newpage
\subsection*{Solid--Liquid Friction}

The friction coefficient $\lambda$ between the solid and the liquid governs the dynamics of interfacial slip and is directly relevant to nanofluidic transport, making it an important dynamical complement to the structural and spectroscopic observables described above.
We assess friction using two complementary metrics: a scalar comparison of the plateau (long-time) value of $\lambda$, and a function-based comparison of the full Green-Kubo running integral $\lambda(\tau)$, which retains information about the transient approach to the plateau that the scalar value alone discards.
The scalar agreement is defined analogously to the descriptor agreements used for SFG,
\begin{equation}
    d^{\lambda}_{\textrm{scalar}} = 1-\frac{\left|\lambda^{\textrm{XC}}-\lambda^{\textrm{RPA}}\right|}{\left|\lambda^{\textrm{RPA}}\right|+\left|\lambda^{\textrm{XC}}\right|},
\end{equation}
while the function-based agreement uses a norm analogous to that used for the RDFs and VDOS,
\begin{equation}
    d^{\lambda}_{\textrm{func.}} = 1-\frac{\int_{0}^{+\infty} \left|\lambda^{\textrm{RPA}}(\tau)-\lambda^{\textrm{XC}}(\tau)\right|d\tau}{\int_{0}^{+\infty} \left|\lambda^{\textrm{RPA}}(\tau)\right|d\tau + \int_{0}^{+\infty} \left|\lambda^{\textrm{XC}}(\tau)\right|d\tau}.
\end{equation}
The two are averaged to give a single friction score for each XC approximation,
\begin{equation}
    d^{\lambda}_{\textrm{tot.}} = \frac{1}{2}\left(d^{\lambda}_{\textrm{scalar}} + d^{\lambda}_{\textrm{func.}}\right),
\end{equation}
which, after conversion to a percentage, provides the overall friction score.

\newpage
%